\documentclass[nonacm,sigplan]{acmart}

\usepackage{graphicx}
\usepackage{subcaption}

\usepackage[]{hyperref}
\usepackage[normalem]{ulem}
\useunder{\uline}{\ul}{}
\newcommand{\name}[1]{\text{Fastrack{#1}}}
\newcommand{\todo}[1]{
  \noindent 
  \colorbox{yellow} {\scriptsize TODO}
  {\bf[}\textcolor{red}{#1}{\bf]} 
}
\newcommand{\para}[1]{\textbf{#1:}}
\newcommand{\xor}{\oplus}

\begin{document}

\title{\name{}: Fast IO for Secure ML using GPU TEEs}

\author{Yongqin Wang}
\email{yongqin@usc.edu}
\authornote{Both authors contributed equally to this research.}
\affiliation{%
 \institution{University of Southern California}
 \city{Los Angeles}
 \state{CA}
 \country{USA}}

\author{Rachit Rajat}
\email{rrajat@usc.edu}
\authornotemark[1]
\affiliation{%
 \institution{University of Southern California}
 \city{Los Angeles}
 \state{CA}
 \country{USA}}

\author{Jonghyun Lee}
\email{leejongh@usc.edu}
\affiliation{%
 \institution{University of Southern California}
 \city{Los Angeles}
 \state{CA}
 \country{USA}}

 \author{Tingting Tang}
\email{tangting@usc.edu}
\affiliation{%
 \institution{University of Southern California}
 \city{Los Angeles}
 \state{CA}
 \country{USA}}

\author{Murali Annavaram}
\email{annavara@usc.edu}
\affiliation{%
 \institution{University of Southern California}
 \city{Los Angeles}
 \state{CA}
 \country{USA}}

\begin{abstract}
As cloud-based machine learning (ML) applications rapidly expand, ensuring data security during model training and inference has become crucial. Hardware-enforced Trusted Execution Environments (TEEs) offer a secure and practical solution for cloud-based ML. Specifically, GPU-based TEE systems provide both robust security and high performance. These systems consist of a CPU TEE and a GPU TEE, where the CPU TEE manages data movement, and the GPU TEE handles access authentication and computation. The isolation provided by CPU and GPU TEEs enhances security, while the GPU delivers substantial computational resources. However, systems like this suffer significant CPU-to-GPU communication overheads. The sender must encrypt the data and attach a Message Authentication Code (MAC), while the receiver must decrypt and verify the data's integrity. These added computational overheads can increase the communication cost between CPU and GPU TEEs by $12.69-33.53\times$, compared to not using the TEE. 
Depending on ML workloads, these added communication overheads result in GPU TEE inference becoming {$54.12\%-903.9\%$} slower and GPU TEE training {$10\%-455\%$} slower compared to inference and training without GPU TEEs. This added latency is undesirable, as it undermines the high-performance advantages of GPU TEEs, especially in latency-sensitive ML applications.

In this paper, we analyze the specific protocols used for CPU and GPU TEE communication in the recently released Nvidia H100 TEE implementation, also referred to as GPU confidential computing,  and identify three key overheads: 1) CPU re-encryption, where CPU TEEs redundantly re-encrypt data that are already encrypted, 2) limited parallelism in authentication, as the authentication kernel is highly sequential, and 3) unnecessary operation serialization. 

Based on these observations, we propose \name{}, a suite of optimizations that includes 1) a direct communication channel to GPU TEEs, 2) increased parallelism in the authentication kernel through multi-chaining authentication, and 3) a decryption scheme that overlaps PCI-e transmission with the encryption and authentication kernels. With these optimizations, we can significantly reduce CPU-to-GPU communication costs and reduce the end-to-end ML inference/training runtime by up to $84.6\%$ compared to baseline GPU TEE systems. Compared to GPU systems without TEEs, \name{} introduces minimal overhead and, in some cases, matches the speed of non-TEE GPU systems.
\end{abstract}

\maketitle
\pagestyle{plain} % should come right after \maketitle

\section{Introduction}
\label{sec:introduction}

In recent years, machine learning (ML) systems have been adopted in a variety of tasks, from image classification to graph network analysis.  ML providers increasingly leverage cloud computing infrastructures for both model training and inference. However, the training and inference data is often proprietary or sensitive, and when it is sent to cloud servers, they are exposed to a broad spectrum of threats. These include risks from compromised operating systems, physical snooping, side channels, and various other attacks~\cite{snoop1, hashemi2022data}. 

To strengthen security and trust in cloud environments, various hardware-enforced Trusted Execution Environments (TEEs) have been introduced. Examples of CPU-based TEEs include Intel SGX~\cite{sgx}, Intel TDX~\cite{tdx}, Arm TrusZone~\cite{trustzon}, and AMD SEV-SNP~\cite{amdcc}, all of which provide hardware-enforced program isolation (an enclave). These TEEs are designed to prevent unauthorized access to guest virtual machines' memory (even from the host OS) by controlling virtual memory translation hardware. Additionally, they offer protection against certain physical attacks, such as memory bus probing using DRAM encryption and authentication.
While these security features make CPU TEEs secure against some threat models, much of the ML inference and training prefer GPUs to exploit their parallel hardware. 
%struggle with limited parallelism, leading to significant throughput reductions for ML inference and training~\cite{darkNight}. 
To address this desire for ML workloads to run on GPUs, various solutions have been proposed to extend TEE capabilities to commodity GPUs. Notable GPU TEE solutions include Graviton~\cite{graviton}, HIX~\cite{hix}, and Nvidia confidential computing~\cite{nvdiacc}. Nvidia confidential computing is now commercially available and is being offered as a practical solution for secure ML. In this work we will analyze and improve on the H100 Nvidia GPU TEE.     
%solution, we specifically refer to it when discussing GPU TEEs. 

\begin{figure}
    \centering
    \includegraphics[width=8cm]{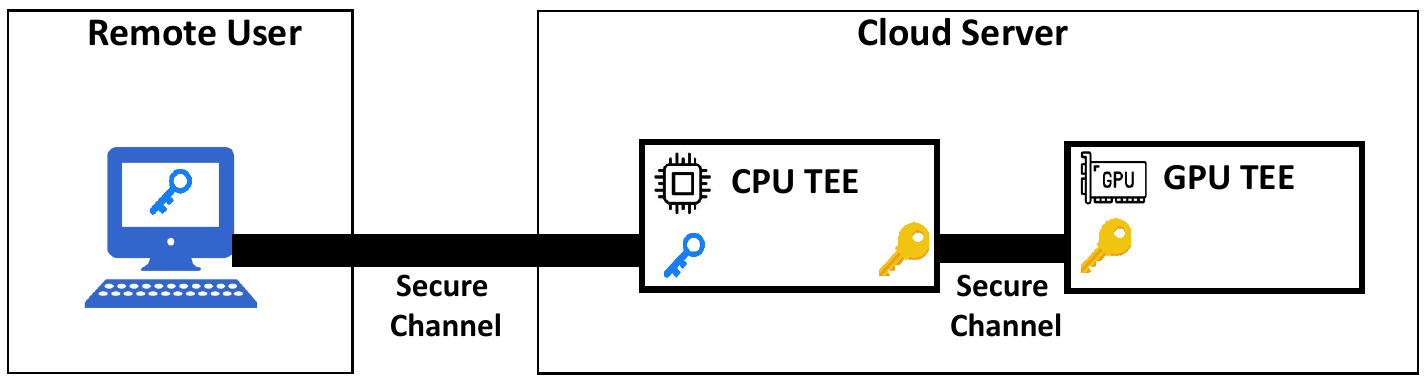}
    \caption{GPU confidential computing system topology.}
    \label{fig:gputee}
\end{figure}

\para{GPU TEE system topology} Figure~\ref{fig:gputee} shows a general system topology for secure ML execution (training/inference) using GPU confidential computing. A GPU's confidential computing (CC) system consists of two major components: 1) CPU TEE and 2) GPU TEE. In the commercially available H100 GPU CC, an Intel TDX-enabled or AMD TEE CPU offers the TEE capabilities needed for hosting GPU drivers and securing CPU-GPU communication. The GPU TEE ensures isolation for computations on the GPU. In this system, the CPU TEE loads the GPU driver as a secure enclave and is responsible for handling the data movements. CPU TEEs will establish secure communication channels with both GPUs and remote users. CPU TEEs can then securely transmit data between remote users and GPUs. The GPU TEEs are responsible for performing data decryption, authenticating data, and then performing the necessary computations. The authentication ensures that the data sent is not tampered with in the presence of untrusted system administrators and physical snooping adversaries on the PCIe path (the precise threat model described later).

With this system, adversaries cannot access or tamper with data/code inside CPU and GPU TEEs. Moreover,  TEEs can provide an attestation method for remote users to verify the code running inside GPUs and CPUs, such that remote users can ensure that their inputs will not be maliciously written out of the TEEs. With those guarantees, model and data owners can entrust their private data to GPU TEE systems, and their data/model will not be revealed to unauthorized users.

\begin{figure*}[htbp]
    \centering
    % Upper subfigure (a)
    \begin{subfigure}{\columnwidth}
        \centering
        \includegraphics[width=\columnwidth]{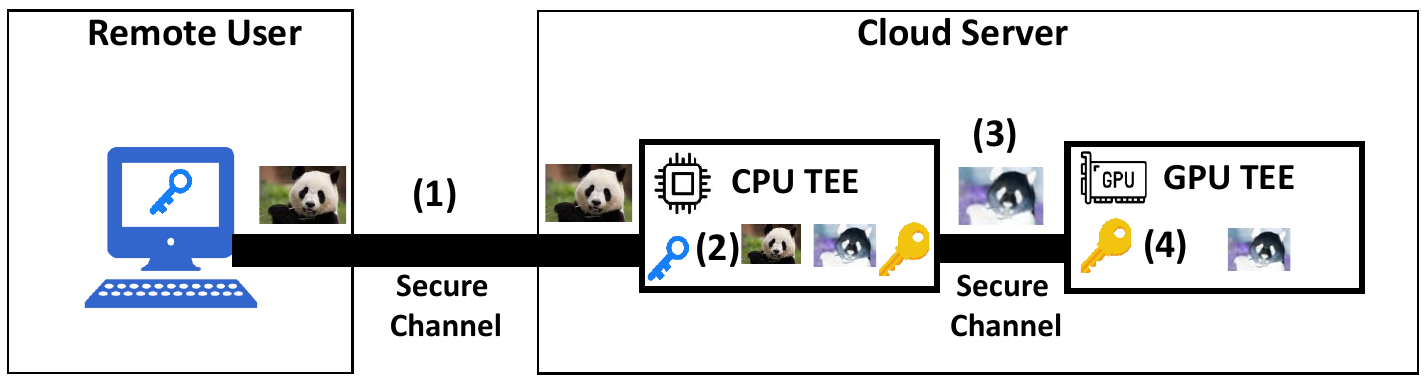}
        \caption{Inference}
        \label{fig:infer-flow}
    \end{subfigure}
    % Lower subfigure (b)
    \begin{subfigure}{\columnwidth}
        \centering
        \includegraphics[width=\columnwidth]{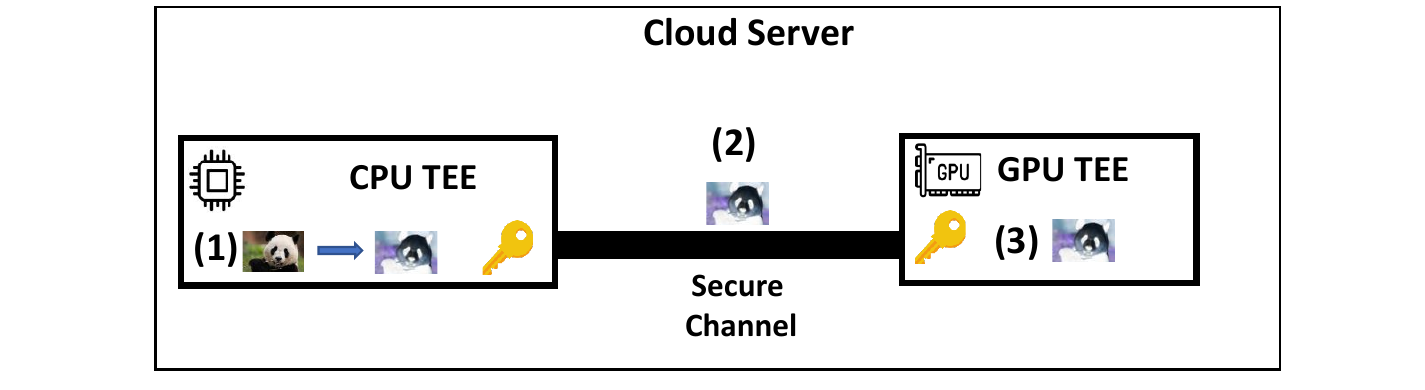}
        \caption{Training}
        \label{fig:train-flow}
    \end{subfigure}
    \caption{ML workflow with GPU TEEs}
    \label{fig:overall_figure}
\end{figure*}

\subsection{ML system using GPU TEEs}
This GPU TEE system can be used to both serve and train ML models. Figure~\ref{fig:infer-flow} and Figure~\ref{fig:train-flow} illustrate how this GPU TEE system is used for inference and training. 

\para{Inference} In Figure~\ref{fig:infer-flow}, model weights will be sent to the GPU enclave before the ML inference services begin. Remote users can use this inference service after the weight has been sent to GPU TEE. This inference process involves three steps: 
\begin{enumerate}
    \item The remote user securely sends data to the CPU TEE.
    \item The CPU TEE performs optional preprocessing on inputs.
    \item The CPU TEE securely sends the data received to the GPU TEE.
    \item The GPU TEE performs inference.
\end{enumerate}

\para{Training} Figure~\ref{fig:train-flow} illustrated the workflow for training. As is common in training protocols,  the training dataset will be placed in the cloud server (CPU TEE) before the start of the training. Thus, training will have the following steps:
\begin{enumerate}
    \item The CPU TEE performs optional preprocessing on inputs.
    \item The CPU TEE securely sends the data received to the GPU TEE.
    \item The GPU TEE performs training.
\end{enumerate}

% \para{Overhead analysis} In the inference process, GPU TEEs introduce additional overheads, specifically in securely transmitting data (Step 3) and performing ownership checks (Step 4). Similarly, during the training process, overheads arise from secure data transmission in Step 2 and ownership checks in Step 5. 
% When securely sending data from party A to party B, party A must encrypt the data and generate a MAC tag on the encrypted data. Party A will also need to transmit encrypted data and the associated MAC tag. After receiving the data from party A, party B needs to decrypt and regenerate a MAC tag to verify against the received MAC tag.

% According to our experiments, secure data transmission can add significant overhead to inference and training processes. \todo{add some examples here.} Those overheads most contributed to added computations to encrypt, decrypt, and authenticate data.

\para{Overheads:} TEE-based ML suffers significant overheads compared to executing these tasks without a TEE. Specifically,  during steps 1 and 3 of inference, there is a significant cost in securely transmitting data between the remote users and CPU TEE and between CPU TEE and GPU TEE. Similarly, during the training process, overheads arise from secure data transmission in Step 2. These overheads arise from the fact that communication channels are outside of the trusted boundaries of TEE, which requires data to be encrypted. Additionally, the data that is placed on the communication channels may also be tampered with by an adversary, compromising the integrity. Hence, generating a MAC tag and transmitting it along with the encrypted data for integrity verification is necessary. When securely sending data from Party A to Party B, Party A must encrypt the data, generate a MAC tag, and transmit both the encrypted data and MAC tag. Upon receiving the data, Party B needs to decrypt it and regenerate a MAC tag to verify against the received tag.

%For the above described steps, the following overheads are expected. During the inference process, GPU TEEs introduce additional overheads, 

We conducted studies to study these secure transmission overheads.
On an H100 Nvidia GPU with CC, CPU-to-GPU communication for even a computationally intensive ResNet increases by 12.69$\times$, resulting in a 54.13$\%$ increase in overall inference time and a 10$\%$ increase in training time. For I/O-intensive ML models, such as Two-Tower Neural Networks (TTNN), the runtime overhead for end-to-end inference can reach 629.9$\%$, and for training, the overhead is 238.9$\%$.
%
% Our analysis showed that the
%These overheads are primarily due to the extra computations needed and sub-optimal implementation for encryption, decryption, and data authentication. We characterize those costs and inefficiencies into three categories and propose multiple optimizations to those additional overheads, which we will elaborate on in the next section.

\subsection{Key observations}
We analyzed Nvidia's GPU CC documentation and created custom kernels to measure the various costs of CC implementation. The following  three are the primary sources of inefficiency in existing GPU CC: 
%To mitigate the added overheads in GPU TEE systems, we introduce a series of optimizations, called \name{}, aimed at minimizing the performance impact of secure communication in GPU TEEs. Our optimizations are based on several key observations about current GPU TEE implementations: 1) CPU re-encryption, 2) limited parallelism for data authentication, and 3) operation serialization.

\para{CPU re-encryption}
During secure ML inference, traditionally the client already encrypts and authenticates the data before sending it to the CPU TEE in the cloud. The CPU TEEs decrypt this data, re-encrypt it using the GPU TEE keys, and generate another MAC tag before sending the data to GPU TEE. This decryption, followed by re-encryption (Steps 2 and 3), introduces overheads. 
Similar inefficiencies arise during training: when datasets arrive at the cloud server, they are already encrypted and have a MAC tag attached. For example, each image in the Imagenet dataset is encrypted and tagged by the client before sending the dataset to the CPU TEE.  The CPU TEE decrypts data and verifies the tag, and then the CPU TEE re-encrypts and regenerates a new tag using the GPU TEE key. The re-encrypt and tag generation are needed here because the PCI-e interface is not trusted. This two-step decryption+authentication followed by re-encryption+MAC generation is expensive. We observe that if the remote users have a shared key with GPU TEEs, they can encrypt and authenticate the data with that key rather than the shared key with CPU TEEs. It’s important to note that the shared key with the GPU TEE is not a secret to the remote user, as this key is already known to the CPU TEE, which the client controls. Thus, without security compromise, the data sent to CPU TEEs can be encrypted and attached with MAC tags using the shared keys with GPU TEEs. Hence, CPU TEEs can then directly deliver the encrypted data and MAC tags to the GPU TEE. Once the data arrives in GPUs, GPU TEEs can proceed with decryption and authentication before utilizing it, streamlining the process.

%To address this inefficiency, we propose a system that allow a direct communication channel with the GPU TEEs. In this approach, once data arrives at the cloud server, the CPU TEEs simply forward the encrypted and authenticated data directly to the GPU TEEs as needed (via CPU TEE but without the penalty for re-encryption and MAC tag generation). 

\para{Lack of parallelism for authentication}
Nvidia's GPU CC documentation states that AES-GCM~\cite{aesgcm} is the encryption and authentication scheme for data transmission among remote users, CPU TEEs, and GPU TEEs~\cite{graviton, nvdiacc}. AES-GCM computes a 128-bit MAC tag over the encrypted data (details on AES-GCM are provided in Section~\ref{subsec:aesgcm}). This tag generation is inherently sequential. Hence, even though a GPU has highly parallel execution resources, the sequential nature of the current AES-GCM is not conducive to exploiting those resources. To address this, we propose a highly parallel authentication approach that breaks large inputs into smaller chunks, allowing for parallel tag generation across multiple input segments and thereby improving throughput.

\para{Operation serialization}
In current implementations, GPU TEEs begin decryption and authentication only after receiving all the transmitted data, such as a batch of images for inference. However, many computations for decryption, such as AES evaluation (more in Section~\ref{subsec:aesgcm}), are input-independent. Those input-independent computations can be performed concurrently while receiving the transmitted data from CPU TEE. Another operation serialization is that the GPU currently waits for authentication to complete before initiating inference and training. We observe that GPUs do not need to wait for authentication to complete; instead, they can eagerly process decrypted data while authentication runs in parallel with data transmission for the next batch. The GPU can initiate computations but buffers all its gradients (for training) or notify remote users that the results are only tentative. In case of authentication failure, the GPU simply abandons these buffered computations or notifies remote users that previous computation results are corrupted.   Based on these observations, we propose a novel GPU decryption pipeline that enables AES evaluation (for the current batch) and authentication (for the previous batch) to occur in parallel with data transfer over the PCI-e bus for the current batch.

\para{Our contributions} We propose \name{} as a collection of three techniques to improve GPU TEE performance and efficiency in secure ML applications. First, a revised communication protocol enables direct communication between the client and GPU TEE, eliminating decryption-encryption costs on the CPU TEE. Second, a GPU-friendly data-parallel multi-chaining authentication scheme optimizes AES-GCM tag computations. Third, a GPU decryption pipeline runs decryption and authentication in parallel with PCI-e data transmission. Together, these optimizations reduce CPU-GPU communication costs, cutting ML inference/training runtime by up to {84.6\%}, with minimal overhead compared to non-TEE GPU systems.

\begin{comment}
We propose \name{} as a collection of three techniques described above that collectively enhance the performance and efficiency of GPU TEEs in secure ML applications. Our contributions are as follows:
\begin{itemize}
    \item An updated communication protocol that enables direct communication between the client and GPU TEE, thereby eliminating the decryption-encryption costs on CPU TEE.
    \item A GPU-friendly data parallel multi-chaining authentication scheme that implements the AES-GCM based tag computations more efficiently.
    \item A GPU decryption pipeline that allows decryption and authentication to run in parallel with data transmission over PCI-e bus.
    \item With those optimization combined, \name{} significantly reduce CPU-to-GPU communication costs and reduce the end-to-end ML inference/training runtime by \todo{up to $84.6\%$} compared to baseline GPU TEE systems.
    \item Compared to GPU systems without TEEs, \name{} introduces minimal overhead and, in some cases, matches the speed of non-TEE GPU systems.
\end{itemize}

\end{comment}

\section{Threat model}
\label{sec:threat}
We assume a threat model identical to that used by the Nvidia confidential computing framework~\cite{nvdiacc}, where the adversary controls the entire system software, including the host operating system, and possesses physical access to the cloud server.

Under this model, the adversary can tamper with or access any process running on the cloud server, as well as data within DMA buffers. The CPU TEE and GPU TEE communication occur via DMA, writing data to a shared bounce buffer (see Section~\ref{subsec:tee} for details). With physical access, the adversary can also carry out bus snooping attacks on PCI-e and DRAM buses connecting the CPU to DRAM.

We trust the integrity of the CPU and GPU packages, meaning we assume the data and code within CPU and GPU TEEs, including confidential virtual machines (CVM) and GPU drivers in CVM, are secure. Like Nvidia confidential computing, we also trust on-package GPU memory, which is closely integrated into the package and considered resistant to external tampering. Identical to Nvidia's confidential computing, side-channel attacks are considered out of scope.

\section{Background}
\label{sec:background}
\subsection{Trusted Execution Environment}
\label{subsec:tee}
Trusted Execution Environments (TEEs) are a security component integrated into modern CPUs. The primary goal of a TEE is to provide a secure, isolated execution environment that guarantees both confidentiality and integrity of sensitive operations. By isolating sensitive processes from the rest of the system, including the operating system, hypervisor, and potential malware, TEEs prevent unauthorized observation or tampering with data. Popular TEE technologies include Intel SGX (Software Guard Extensions), ARM TrustZone, AMD SEV (Secure Encrypted Virtualization), and Intel TDX (Trusted Domain Extensions). In this work, we focus on Intel TDX, a CPU TEE technology that provides isolation capabilities for virtual machines (VMs) from the underlying system, including the hypervisor and other VMs.

\para{GPU-TEEs}
Recent innovations, such as Graviton\cite{graviton} and HIX (Heterogeneous Isolated Execution)~\cite{hix}, have proposed the concept of GPU-TEEs, expanding the TEE model to handle the growing demand for secure, high-performance computing. Notably, Nvidia has released its confidential computing framework as part of its Hopper architecture, marking the first commercial implementation of a GPU-TEE.

The concept of a GPU-TEE is similar to a CPU-TEE, but with the GPU now included as part of the Trusted Computing Base (TCB). Ensuring secure communication between the CPU-TEE and the GPU-TEE is critical. To achieve this, the GPU driver is placed inside the CPU-TEE, preventing any adversary from directly accessing the GPU. 
Additionally, data transferred between the CPU-TEE and GPU-TEE is encrypted and authenticated using AES-GCM(Section~\ref{subsec:aesgcm}), ensuring that the data remains confidential and unmodified during transmission on untrusted PCIe channels. CPU TEEs need to encrypt and generate a MAC tag to send data to GPU TEEs.
After encryption and MAC tag generation, CPU TEEs will put data encrypted inside a bounce buffer. The bounce buffer is a memory region from which GPU TEEs can directly read (through PCI-e). After GPU TEEs read the encrypted data into its memory, it decrypts/authenticates the data. 

%While encrypting and authenticating data between the CPU-TEE and GPU-TEE ensures security, it adds extra overhead to CPU-to-GPU communication. Extra computation is needed for encryption and authentication on both the CPU and GPU side. Optimization of this process is the key contribution of our paper.

\subsection{AES-GCM}
\label{subsec:aesgcm}

AES-GCM (Advanced Encryption Standard - Galois/Counter Mode) is a popular cryptographic algorithm that provides both confidentiality (encryption/decryption) and integrity (authentication) in an extremely efficient fashion. AES-GCM combines the confidentiality guarantees of the AES (Advanced Encryption Standard) block cipher with the integrity guarantees of the GMAC, making it a popular choice for implementing secure communication in TEEs~\cite{graviton,nvdiacc}. 

%suitable for various applications, like data protection, secure communications, and more. 

\para{The AES block cipher} 
AES (Advanced Encryption Standard) is a symmetric block cipher that encrypts data in fixed-size 128-bit blocks. It uses a secret key $k$, typically 128, 192, or 256 bits, to transform a plaintext block through multiple rounds of permutations. Each round applies key-dependent transformations like SubBytes, ShiftRows, and MixColumns to create ciphertext. AES is widely used for secure data encryption due to its efficiency and strong resistance to cryptographic attacks.

\begin{figure}[h]
    \centering
    \includegraphics[width=6cm]{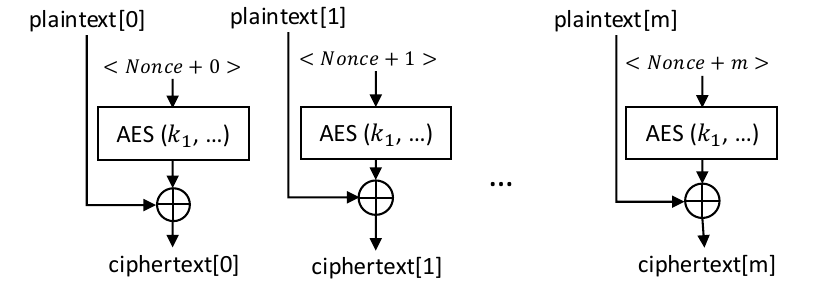}
    \caption{AES-CTR mode to encrypt $m$ blocks of plaintext.}
    \label{fig:aecctr}
\end{figure}

In AES-GCM, AES is used in counter mode (CTR). Figure~\ref{fig:aecctr} illustrates how AES-CTR uses AES to construct a stream cipher. To encrypt a message of $m$ blocks, AES-CTR starts by selecting an initial nonce (a unique, random value) and a counter for the encryption process. This combination forms the first counter block. AES encrypts each counter block using the secret key $k_{1}$, producing a block of random bits. The resulting random-bit block is then $\xor$-ed with a block of plaintext data to produce the corresponding ciphertext block. Decryption follows the same procedure: the counter blocks are encrypted with AES to produce the same random-bit blocks used for encryption, which are then $\xor$-ed with the ciphertext blocks to retrieve the original plaintext. This AES-CTR is highly parallel as each AES block evaluation is independent of other AES blocks.

\para{Authetication with Galois Field} AES-GCM computes an authentication tag using Galois Field arithmetic over GF($2^{128}$) to ensure the integrity of both the ciphertext and any associated data (AAD). The tag verifies that neither the encrypted data nor the AAD has been tampered with. Inputs for authentication include ciphertext from AES-CTR mode, AAD for integrity protection (e.g., headers), a unique session nonce, and the lengths of both ciphertext and AAD. A hash subkey $H$, derived by encrypting an all-zero block with the encryption key, is used for Galois Field multiplication during authentication.
\begin{comment}
In addition to encryption, AES-GCM computes a message authentication code (MAC) called \textbf{authentication tag} using Galois Field arithmetic over GF($2^{128}$) to ensure that the ciphertext and any additional associated data (AAD) have not been tampered with. The authentication tag is computed over both the ciphertext and the AAD, offering integrity and authenticity checks. %for both encrypted and non-encrypted data.

The input data required for authentication are:
\begin{itemize}
    \item Ciphertext: The encrypted data, produced by AES-CTR mode
    \item Additional Authenticated Data (AAD): Non-encrypted data that need integrity protection, such as protocol headers or metadata.
    \item Nonce: A unique value generated for each encrypted session. This value is critical for ensuring security and is included in the tag computation
    \item Message Length: The lengths of both the ciphertext and AAD are also included in the tag computation
\end{itemize}
For authentication, a special value, called the hash subkey $H$, is derived by encrypting an all-zero block with AES using the encryption key. This subkey $H$ is used throughout the authentication process for Galois Field multiplication. 
\end{comment}

\begin{figure}[h]
    \centering
    \includegraphics[width=6cm]{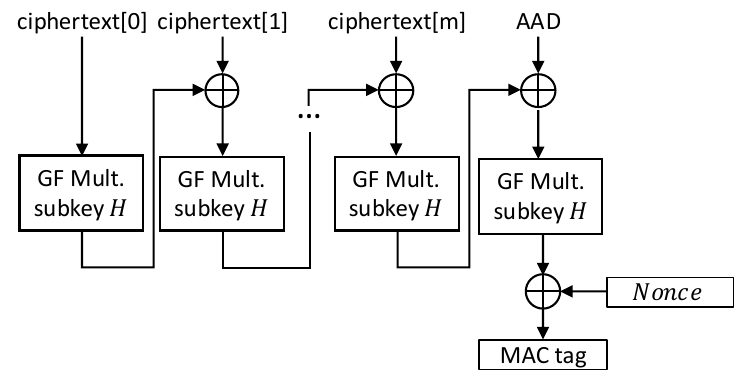}
    \caption{GMAC used in AES-GCM MAC tag generation.}
    \label{fig:gmac}
\end{figure}

Figure~\ref{fig:gmac}shows the GMAC tag computation, where ciphertext and AAD are divided into 128-bit blocks, with padding for smaller blocks. Each block (except the first) is $\xor$-ed with the previous block's product with subkey $H$, then multiplied with $H$ using Galois Field arithmetic. This continues block by block until all are processed. The final product is $\xor$-ed with the nonce to produce the MAC tag. Despite GPU parallelism, AES-GCM authentication remains inherently sequential, limiting parallel processing.

 % This poses a significant challenge to CPU/GPU TEE communication. 
%YR: WE CAN COMMENT THE BELOW PART IF WE ARE SHORT ON SPACE. 
\begin{comment}
To address this challenge, our work introduces a method for increasing parallelism within the authentication process. Specifically, we propose breaking down the serial operations into smaller, independent chains, allowing multiple authentication computations to proceed concurrently. This approach reduces bottlenecks and accelerates the overall authentication process, making it more efficient for CPU/GPU TEE communication.
\end{comment}

\section{Motivational Data for Fastrack}
\label{sec:pipe}
\label{subsec:fbaseline}

\begin{figure}[h]
    \centering
    % Upper subfigure (a)
    \begin{subfigure}{\linewidth}
        \centering
        \includegraphics[width=\columnwidth]{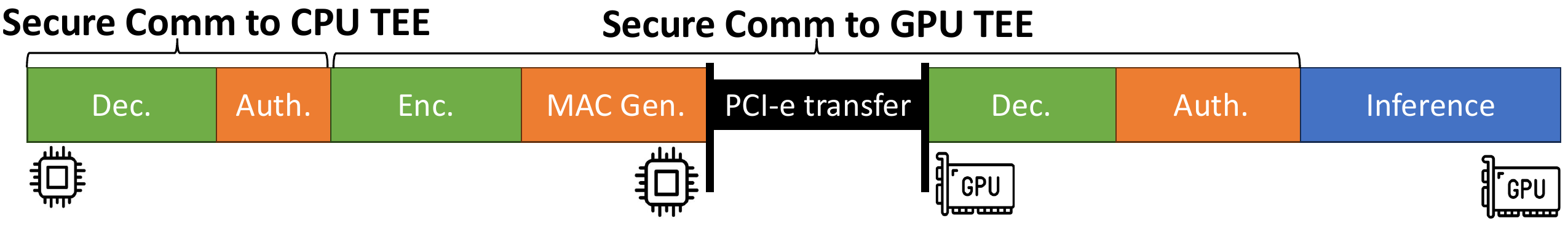}
        \caption{Inference.}
        \vspace{0.5cm}
        \label{fig:infer-compute}
    \end{subfigure}
    % \vspace{1cm}
    % Lower subfigure (b)
    \begin{subfigure}{\linewidth}
        \centering
        \includegraphics[width=\columnwidth]{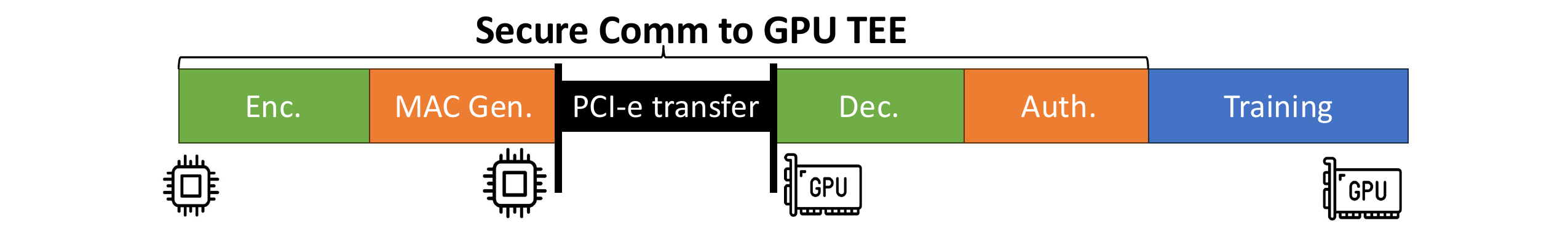}
        \caption{Training.}
        \label{fig:train-compute}
    \end{subfigure}
    \caption{Baseline computation and communication flow.}
    \label{fig:baseline-compute}
\end{figure}
We begin by providing quantitative motivational data for the baseline GPU CC system. In Figure~\ref{fig:gputee}, we outline the steps for ML inference and training within GPU CC. Figure~\ref{fig:baseline-compute} provides a detailed breakdown of the computations and communications performed in CPU and GPU TEE within the CC framework. For clarity, we omit any optional preprocessing steps in Figure~\ref{fig:baseline-compute}.

\para{Inference computation flow} A batch of inference data is encrypted using the shared key between CPU TEEs and the remote user (the blue key in Figure~\ref{fig:gputee}). Since the CPU TEE manages data movement, it will first decrypt (labeled as \textit{Dec} in Figure~\ref{fig:infer-compute}) and authenticate (labeled as \textit{Auth}) the data to check for any errors introduced by adversaries. Once authentication is successful, the CPU TEE uses the GPU driver (which shares a key with the GPU TEEs) to securely transfer the data to the GPU TEEs. Due to the potential access adversaries may have to the DMA bounce buffers and PCI-e buses, an additional round of encryption (\textit{Enc}) and MAC generation (\textit{MAC Gen}) is performed. After receiving the data, the GPU TEE will decrypt and authenticate it. If this authentication is successful, the GPU TEE will proceed with running inferences on the decrypted data. This process is repeated for each inference batch.

\para{Training computation flow} Figure~\ref{fig:train-compute} shows the  flow for training. Typically before secure training, the training data is already stored inside CPU TEEs.
CPU TEE then randomizes the training data and selects one training batch from the dataset to deliver them to the GPU TEEs. 
This process will involve encryption, MAC generation on CPU TEE, and decryption and authentication on GPU TEE. Once authentication is complete, the GPU TEE proceeds with training on the decrypted data. Note that if the client transmits training data per batch rather than sending all the data a priori, the training pipeline will be identical to the inference pipeline, which brings the CPU decryption and authentication times into the training path.   

%Before the training process, we assume that all the datasets are already available inside CPU TEEs and the model weights are initialized inside GPU TEEs. Additionally, we assume that the granularity of dataset authentication is at the level of a single input sample. For instance, in the case of ImageNet, each image is associated with one MAC tag. During the training process, the dataset is already available inside CPU TEEs. Thus, CPU TEEs do not need to decrypt data. 

\begin{table}[]

\caption{Data transfer time per batch for CC vs non-CC versions.}

\begin{tabular}{|c|c|c|c|}
\hline
{\ul \textbf{Model}}                                                           & \textbf{\begin{tabular}[c]{@{}c@{}}No GPU \\ TEE\\ (Train)\end{tabular}} & \textbf{\begin{tabular}[c]{@{}c@{}}NVIDIA\\ CC \\ (Train)\end{tabular}} & \textbf{\begin{tabular}[c]{@{}c@{}}NVIDIA\\ CC \\ (Inf.)\end{tabular}} \\ \hline
\textbf{\begin{tabular}[c]{@{}c@{}}Resnet50\end{tabular}}              & 312.5$\mu$$s$                                                                     & 3.96ms                                                                  & 6.41ms                                                                   \\ \hline
\textbf{GraphSAGE}                                                             & 2.1ms                                                                    & 32.6ms                                                                 & 45.5ms                                                                 \\ \hline
\textbf{\begin{tabular}[c]{@{}c@{}}TTNN\end{tabular}} &         2.19ms                                                             & 51.43ms                                                                   & 73.45ms                                                                  \\ \hline

\end{tabular}
\label{table:motivation1}
\end{table}

\begin{table}[]

\caption{Total execution time per batch for CC vs non-CC versions.}

\begin{tabular}{|c|c|c|c|c|}
\hline
{\ul \textbf{Model}}                                                           & \textbf{\begin{tabular}[c]{@{}c@{}}No GPU \\ TEE\\ (Train)\end{tabular}} & \textbf{\begin{tabular}[c]{@{}c@{}}NVIDIA\\ CC \\ (Train)\end{tabular}} & \textbf{\begin{tabular}[c]{@{}c@{}}No GPU \\ TEE\\ (Inf.)\end{tabular}} & \textbf{\begin{tabular}[c]{@{}c@{}}NVIDIA\\ CC \\ (Inf.)\end{tabular}} \\ \hline
\textbf{\begin{tabular}[c]{@{}c@{}}Resnet50\end{tabular}}              & 35.84ms                                                                   & 39.49ms                                                                  & 11.27ms                                                                   & 17.37ms                                                                \\ \hline
\textbf{GraphSAGE}                                                             & 8.66ms                                                                    & 39.16ms                                                                    & 5.41ms                                                                   & 48.9ms                                                                 \\ \hline
\textbf{\begin{tabular}[c]{@{}c@{}} TTNN\end{tabular}} & 35.55ms                                                                    & 84.96ms                                                                   & 13.46ms                                                                   & 84.79ms                                                                  \\ \hline
\end{tabular}
\label{table:motivation2}
\end{table}
\para{Motivational results} We implemented three ML models to showcase the overheads of GPU TEE systems. Those models include Resnet50~\cite{resnet}, GraphSAGE~\cite{gcn}, and Tower Tower neural network~\cite{ttnn1, biasttnnpaper} (details about those models will be in Section~\ref{sec:evaluation}). Table~\ref{table:motivation1} shows the transfer runtime (in milliseconds) per batch for the three ML models mentioned with and without Nvidia CC. 

Nvidia's GPU CC documentation~\cite{nvdiacc} describes the security protocols but does not present the implementation details.  Since we are proposing to make modifications to the GPU CC, we implemented the baseline security protocol following the description in the Nvidia documentation using OpenSSL library \cite{openssl} for CPU encryption and MAC generation, while GPU decryption and authentication were handled via custom CUDA kernels. In this section, the motivation data is derived from our implementation. However, since we can measure the end-to-end runtime on actual CC systems, in the evaluation section, we will compare the performance of our scheme directly against the Nvidia CC systems. 

The transfer time refers to the cumulative time across all the steps shown in Figure~\ref{fig:baseline-compute}, except the last component of inference or training computation.  
As can be seen, for the benchmarks mentioned, there is a $12.69-23.48\times$ slowdown for data transfer when using our implemented baseline for training and up to $33.53\times$ slowdown for inference.
The larger slowdown during inference is primarily due to the extra decryption and authentication that CPU TEEs must perform because data from remote users arrives in real-time. 
%
%This additional decryption and authentication process by the CPU TEE is relatively slow, contributing to the greater inference communication overheads.

Table \ref{table:motivation2} highlights the end-to-end inference and training slowdowns due to these transfer time overheads. Among the three models, GraphSAGE GCN experienced the highest percentage increase in runtime. GCNs generally only have a couple of convolution layers, and hence, the GPUs perform these computations quickly and end up waiting for the data delivery from CPUs. With GPU CC, data transfer times accounted for up to 91.93$\%$ and 93.04$\%$ of the total runtime for training and inference, respectively. While one might argue for running GraphSAGE on the CPU alone, our observations showed that the CPU-only execution was still slower than running it on a GPU with confidential computing enabled. Thus, while the GPU CC is expensive, running training only on the CPU is still not an attractive choice for designers. Similar trends were observed with TTNN. We observe that training ResNet50 is compute-intensive since there are dozens of convolution layers requiring significant matrix computations. Thus, for each batch of data that is delivered to the GPU, the GPU computes for a larger fraction of the total time,  resulting in a relatively modest slowdown of around 10$\%$ when using CC.

\section{Fastrack Design}
As demonstrated by the motivational data, ML workloads in GPU TEE systems face redundancies and sub-optimal implementations, leading to significant overhead. In the following sections, we explore these inefficiencies and present our proposed solutions to address them. We present our three optimization techniques in Sections~\ref{subsec:fcpuremove},~\ref{subsec:fmultichain}, and~\ref{subsec:fdecryptpipe}. Finally, we will discuss~\name{}'s security in Section~\ref{subsec:secanalysis}.

\subsection{Direct communication with GPU TEEs}
\label{subsec:fcpuremove}
In current secure ML implementations (such as the Nvidia's reference implementation~\cite{nvdiacc}) all inputs from users and datasets are sent to CPU TEEs first, and CPU TEEs then send those data to GPU TEEs. During this process, CPU TEEs decrypt and authenticate the user-supplied data and then re-encrypt and generate tags for transmission to GPUs.
For inference, when user inputs arrive at cloud servers, their inputs are already encrypted, and a MAC tag is attached. Because the GPU TEEs do not have a shared key with remote users, it is the CPU TEE's job to decrypt and re-encrypt received data with the shared keys with the GPU TEEs.
This two-step process is inefficient because the CPU's encryption and authentication are relatively slow. Moreover, from a security perspective, encrypted inputs from the user and re-encrypted data from the CPUs have the same security guarantees since they all use AES-256-GCM. The only reason that CPU TEEs are doing the re-encryption process is that GPU TEEs do not know the key of remote users, as seen in Figure~\ref{fig:gputee} (GPU TEEs do not have the blue key). This redundancy also exists for training, as the dataset is encrypted with a key unknown to the GPU. As a result, CPU TEEs must re-encrypt the data and generate MAC tags for the batched training inputs before sending them to the GPU TEEs.

\begin{figure}[h]
    \centering
    \includegraphics[width=8cm]{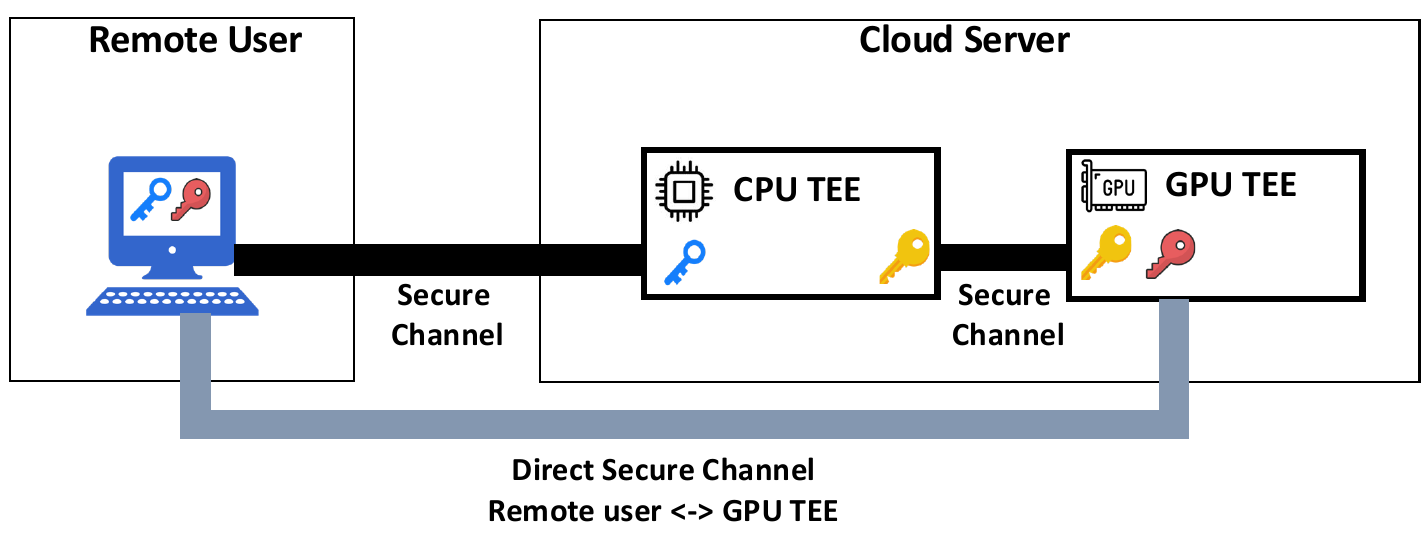}
    \caption{GPU TEE system with shared user and GPU TEE keys.}
    \label{fig:sharedkeytee}
\end{figure}

\begin{figure}[h]
    \centering
    \includegraphics[width=\columnwidth]{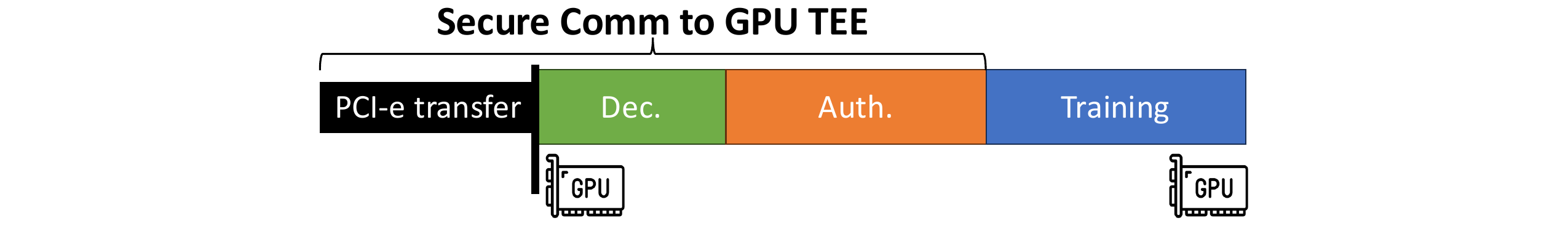}
    \caption{Server computation flow with direct communication with GPU TEEs.}
    \label{fig:sharekey-compute}
\end{figure}

To address this inefficiency, we propose a new system where GPU TEEs share a key with remote users, as shown in Figure~\ref{fig:sharedkeytee}. The remote user will use the Diffie-Hellman key exchange protocol (the same key-sharing protocol used in baseline CPU TEEs and GPU TEEs) to share a symmetric key. This shared key enables remote users to establish a direct secure channel with the GPU TEEs. With this direct and secure channel between remote users and GPU TEEs, remote users can just use the shared key with GPU TEEs to encrypt and generate tags for inputs and datasets and then send them to the CPU TEE in the cloud.
When encrypted inputs/datasets arrive at the remote servers, the CPU TEEs do not need to re-encrypt data before sending it to the GPUs, as the GPU TEEs can handle encryption and authentication directly. In this setup, the CPU TEEs simply place the encrypted inputs and associated MAC tags in a bounce buffer, which is directly transferred to the GPU TEE through PCI-e. As mentioned in Section~\ref{sec:background}, content in this bounce buffer can be read directly by GPU TEEs. With our shared key system, the server computation workflow for inference and training aligns with the process illustrated in Figure~\ref{fig:sharekey-compute}, where CPUs are no longer in the critical path of data transmission to GPU TEEs.

% The Nvidia confidential computing has DMA engines with encrypt/decrypt capability, which are responsible for data movement to and from the CPU’s memory.

% In a confidential environment, DMA engines can only access shared memory pages to retrieve and place data. To ensure the confidentiality and integrity of the payloads, models, and data, the data in these pages are encrypted and signed. These shared memory regions are called bounce buffers because they will be used to stage the secured data before the data is transferred into the secured memory enclaves, decrypted and authenticated, and then processed.

\para{Preprocessing}
One reason CPU TEE may want to decrypt the data is to perform some preprocessing. For example, in image classification systems, CPUs may perform random crop, random horizontal flip, and normalization on input images. In the baseline GPU TEE systems, CPUs would perform preprocessing before sending data to GPU TEEs. 
However, the intention of our shared-key GPU TEE system is to remove CPU TEEs from the critical path of communication to GPUs. In our modified direct communication scheme we propose to use GPU to perform preprocessing. In fact, the recent release of preprocessing libraries on GPUs~\cite{dali} makes GPUs an ideal choice for preprocessing.  After GPUs decrypt data and before inference/training, GPUs will perform preprocessing and then feed the preprocessed images to ML models. This comes with two advantages. The first advantage is the reduced number of bytes transmitted over PCI-e.
For the ImageNet~\cite{imagenet} dataset, preprocessed images are much larger than the original images in JPEG format~\cite{jpeg} (compressed). On average, unprocessed images are around {112KiB} and are around 5$\times$ smaller than the preprocessed images, which is around 588kiB. Another advantage of using GPU to perform processing faster is preprocessing time. Due to more parallelism in the most recent commodity GPUs~\cite{h100,blackwell} and updates in the processing library for GPUs~\cite{dali}, GPU preprocessing is faster than its CPU counterpart. Thus, preprocessing in the GPU TEEs eliminates the need for CPU TEEs to re-encrypt and re-authenticate data. With this new direct communication channel establishment, we now discuss how GPU TEEs handle encrypted data received from the CPU TEE.

\subsection{Multi-chaining authentication}
\label{subsec:fmultichain}
Once GPU TEEs receive the encrypted data, they need to decrypt and authenticate the received data. The detailed hardware and software implementation of the decryption and authentication on   Nvidia's CC has not been made publicly available to the best of our knowledge. %, GPU TEEs use specialized hardware in the DMA engine for decryption and authentication. Due to the lack of access to this hardware, 
Hence, we implemented an optimized AES-GCM kernel for decryption and authentication as a baseline and then implemented our proposed solutions on top of this baseline to show the relative improvements. Because we use AES-GCM to encrypt and authenticate the data, the decryption kernel will be AES-CTR, and the authentication kernel will compute the GMAC using the algorithm described in Section~\ref{subsec:aesgcm}. In our experiments, the decryption kernel is much faster than the authentication, as shown in Figure~\ref{fig:sharekey-compute}. This feature is because GMAC does not have as much parallelism as AES-CTR does. As discussed in Section~\ref{subsec:aesgcm}, AES-CTR divides inputs into $m$ 128-bit blocks and evaluates $m$ AES blocks in parallel, whereas the GMAC function will sequentially evaluate $m$ block multiplications. Due to this sequential chaining process, the GMAC takes longer to compute.

\begin{figure}[h]
    \centering
    \includegraphics[width=6cm]{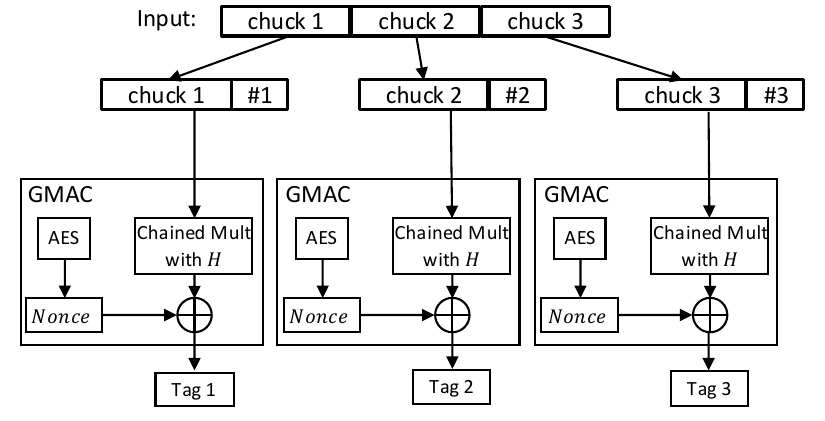}
    \caption{An example of multi-chaining authentication with three chains.}
    \label{fig:3-chain}
\end{figure}

To maximize the speed of our authentication kernel, increasing parallelism is crucial. We propose a multi-chaining authentication approach that enhances parallelism and accelerates the authentication process. Rather than performing a single, continuous chain of block multiplications to generate one MAC tag per input sample, our approach breaks the chain into multiple smaller chunks, resulting in several MAC tags for each input, as shown in Figure \ref{fig:3-chain}. In this figure, a large input is split into three smaller chunks, each attached to one byte of additional authentication data (AAD) to indicate the chunk order (with up to 16 chains, one byte suffices for tracking order). Each chunk, along with its AAD, is processed through the GMAC function to obtain one MAC tag. For example, in the default implementation, every image in a batch has a single MAC tag generated by the client. However, in the multi-chaining approach, a single image may be split into 16 chunks. Each chunk with its own TAG and an AAD is then transmitted to GPU for decryption and verification. This segmentation introduces more parallelism into authentication kernels, allowing it to operate 5-15$\times$ faster.

%YR: COMMENTING THIS FIGURE FOR SPACE
\begin{comment} 
\begin{figure}[h]
    \centering
    \includegraphics[width=\columnwidth]{figures/multi-chain.pdf}
    \caption{Server computation flow with direct communication with GPU TEEs and multi-chaining authentication.}
    \label{fig:multi-chain}
\end{figure}
\end{comment}

\para{Costs of Multi-chaining authentication} Our multi-chaining authentication slightly increases the data transmitted over PCI-e and marginally adds to the number of AES blocks evaluated. However, these increases are inconsequential (less than 1\%) compared to the speedups achieved in the authentication kernels. Assume we use $n$ chains when authenticating one input. Compared to authentication with a single long chain, $n$-chaining authentication requires transmitting $n-1$ additional MAC tags and $n$ extra bytes of AAD. The GPU TEEs will also evaluate $n-1$ more AES blocks. These cost increases result from the greater number of chains; for each chain, we generate one additional MAC tag and evaluate one more AES block for the nonce (see Section~\ref{subsec:aesgcm} and Figure~\ref{fig:3-chain}). Typically, $n$ ranges from 8 to 16, depending on the workload. Compared to the data transmitted for inputs and the number of AES blocks evaluated for encryption, this added overhead is practically insignificant. We will provide concrete figures in Section~\ref{sec:evaluation} to show that those increases are virtually unobservable.

\subsection{Decryption pipeline}
\label{subsec:fdecryptpipe}
With the direct connection with GPU TEEs and multi-chaining, the authentication cost is now reduced. 
% computation workflow for both ML inference and training looks like the one in Figure~\ref{fig:multi-chain}
Still, PCI-e transmission, GPU decryption, and the parallelized authentication kernel remain on the critical path. %At first glance, it seems like there is no further optimization because the decryption and the authentication can only be launched after GPU TEEs receive the encrypted messages and associated MAC tags. However, when 
After analyzing the AES-GCM's algorithm, we find two techniques that allow us to overlap this PCI-e transfer time with the majority of the decryption and authentication kernel computation time. Those two techniques are parallel AES evaluation and eager evaluation. The first technique allows decryption kernels to run in parallel with the PCI-e transfer, and the second technique allows the authentication kernels to run in parallel with the PCI-e transfer as well.

\begin{figure}[h]
    \centering
    \includegraphics[width=\columnwidth]{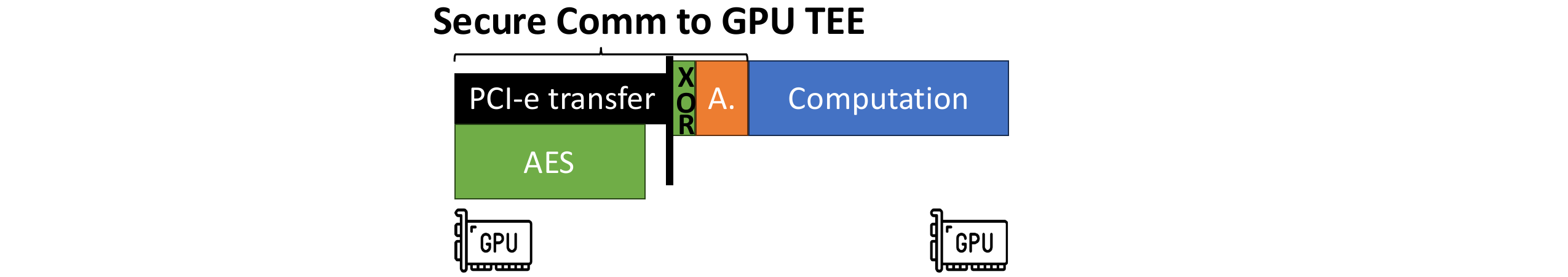}
    \caption{Server computation flow with direct communication with GPU TEEs, multi-chaining authentication, and parallel AES evaluation.}
    \label{fig:compute-optpara}
\end{figure}

\para{Parallel AES evaluation}
The first technique, parallel AES evaluation, allows part of the decryption kernel to run in parallel with the PCI-e transfer. In Section~\ref{subsec:aesgcm}, we have learned that to decrypt a data block $c$ with a key $k$ in AES-GCM, one needs to evaluate this function $AES(k, x)\xor c$, where $x$ is a predefined counter. Note that on the left-hand side, function $AES(k, x)$ is completely independent of the input ciphertext. Its main function is to generate a one-time pad (OTP) using the key, and this OTP generation part is the main bottleneck for decryption. The input decryption task is then just a simple $\xor$ between ciphertext and the OTP generated as AES output. This means that to perform the AES block evaluation, we do not need to wait for the encrypted data block. Thus, we can overlap the OTP generation component of the AES evaluation in the decryption kernel, which makes up most of the decryption kernel, with the PCI-e transfer. The updated computation flow graph is in Figure~\ref{fig:compute-optpara}.

% \para{Eager evaluation} With most of the decryption kernel hidden under the PCI-e, we take steps further to hide the authentication kernel with the PCI-e transfer. To achieve this, we rely on a key observation that after encryption steps, GPU TEEs can eagerly begin computation for inference and training without waiting for the authentication to complete because, after the decryption kernel, the original data block is revealed. After inference/training is done with this batch of data, we can launch the authentication kernel with the PCI-e data transfer for the next batch of data. In this way, we can hide the authentication kernel with PCI-e transfer as well. This overlap is illustrated in Figure~\ref{fig:complete}. According to our experiments in Section~\ref{sec:evaluation}, the PCI-e transmission can hide most of the authentication kernel execution time.

\begin{figure}[h]
    \centering
    \includegraphics[width=\columnwidth]{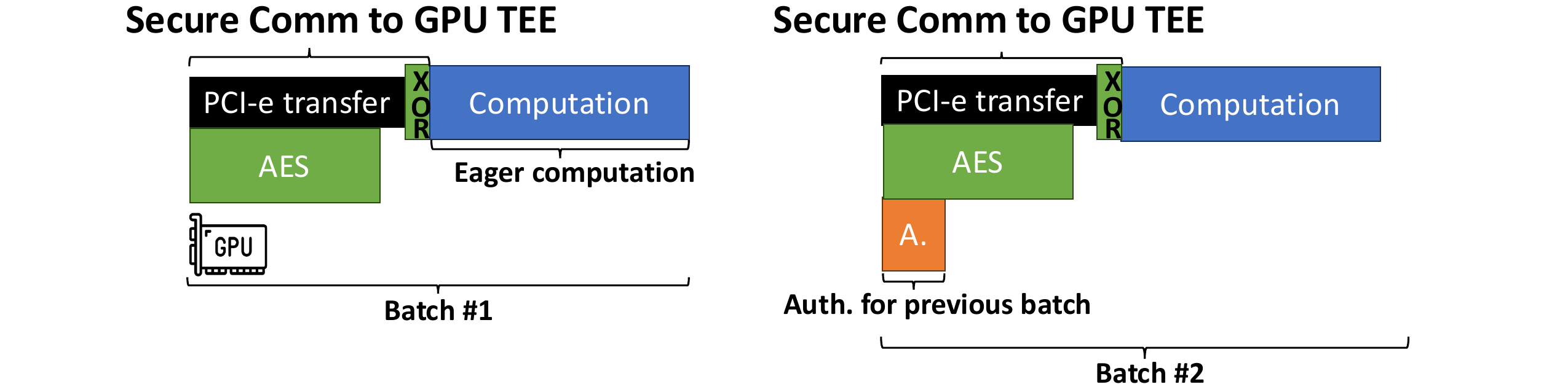}
    \caption{Server computation flow with all optimizations from \name{}.}
    \label{fig:complete}
\end{figure}

\para{Eager evaluation} With much of the decryption kernel already hidden under PCI-e transfer, we take it a step further by overlapping the authentication kernel with PCI-e transfers. We propose a speculative approach where after the decryption kernel, GPU TEEs can eagerly start inference and training computations \textit{without} waiting for authentication to complete, as the original data block is already revealed. Once inference or training on the current data batch starts, the authentication kernel can then run in parallel with the PCI-e transfer for the next batch. This overlap, shown in Figure~\ref{fig:complete}, allows the PCI-e transmission to conceal much of the authentication kernel’s execution time, as we will demonstrate by our experiments in Section~\ref{sec:evaluation}.

\subsection{Security analysis}
\label{subsec:secanalysis}
\name{} provides identical security as the original GPU CC against the threat model described in Section~\ref{sec:threat}.

\para{Confidentiality} Adversaries in our threat model are able to observe the traffic between those parties: \textbf{1) Remote user <-> CPU TEEs, 2) Remote user <-> GPU TEEs, and 3)  CPU TEEs <-> GPU TEEs}. The adversary can observe traffic by probing physical DRAM or PCI-e buses or by directly reading IP packets and DMA buffers. However, this adversary cannot access data/code stored inside CPU/GPU TEEs. In our system, the adversary gains no additional information about the plaintext data within this traffic. All communication between parties is encrypted using AES-GCM with a 256-bit key. For an adversary to compromise confidentiality, they would need to break AES-GCM encryption, which has been demonstrated to be computationally infeasible~\cite{aesgcm}.

\para{Integrity} Adversaries can also modify the traffic among remote users, CPU TEEs, and GPU TEEs. Data transmitted among those parties comes with a MAC tag generated using AES-GCM. It is computationally infeasible for an adversary to introduce the error to transmitted data while passing the MAC tag check. To break the integrity of our system, the adversary will have to break the MAC tag-checking mechanism, which is computationally infeasible.

\para{Security of direct communication with GPU TEEs} Our approach described in \ref{subsec:fcpuremove} does not add any new vulnerabilities. We use the same secure key-exchange protocol that Nvidia-Confidential computing TEEs use, i.e., Diffie Hellman. Instead of a secure data transfer between the remote user and CPU TEE first and then another secure transfer between CPU TEE and GPU TEE, we have a single secure transfer between the remote user and GPU TEE. This is secure due to the security guarantees provided by the Diffie-Hellman key exchange and the AES-256-GCM authenticated encryption scheme. Note that the client would know the GPU TEE key even in the baseline since that key is used by the CPU TEE to encrypt the data sent to the GPU. 

\para{Security of decryption pipeline} Our decryption pipeline mentioned in section \ref{subsec:fdecryptpipe} is secured as here we do not modify the security protocols used. In this approach, we are pipelining the data transfer with the decryption and authentication used. Our approach still uses the same decryption and authentication; it's just pipelined now. Parallel AES evaluation mentioned in section \ref{subsec:fdecryptpipe} simply precomputes the one-time pad used in AES during the data transfer so that when the data is fully transferred, decryption simply requires a simple XOR. Eager evaluation simply postpones the authentication so that data processing can start in advance. The authentication scheme we will use is AES-GCM, whose integrity guarantee is proven secure.

\para{Security of multi-chaining authentication} Our multi-chaining authentication creates multiple tags per input rather than having a single tag. The tag generation used is still the same as used in Nvidia-confidential computing TEEs. This ensures the same integrity as Nvidia-confidential computing TEEs. One might wonder if there can be some reordering attacks due to multi-chaining authentication. We will discuss that in the next paragraph. 
 
\para{Reorder attacks for multi-chaining authentication} For a single input like an image or graph, we generate multiple MAC tags, each tied to a portion of the input. For a large input $X$, it is split into $n$ chunks, $X_1, X_2, ..., X_n$, with each chunk having its own MAC tag, $T_1, T_2, ..., T_n$. Adversaries could reorder these chunks, allowing GPU TEEs to receive, for example, $X_4, X_7, X_8, ..., X_{n-1}$ along with their corresponding tags $T_4, T_7, T_8, ..., T_{n-1}$.

Our system detects reordering attacks by embedding a 1-byte serial number in the additional authenticated area of each chunk (Section~\ref{subsec:aesgcm}), which tracks the order of received chunks. With a maximum value of $n = 16$, 1 byte is sufficient for a single image. For an adversary to tamper with this serial number, they would need to break AES-256-GCM, which is computationally infeasible.  Note that multi-chaining will transfer a few more bytes of data over the baseline. Hence, the adversary may know the presence of multi-chaining (a side channel), but that does not change the security guarantees as the adversary has no information about the data. 

\section{Evaluation}
\label{sec:evaluation}

As stated earlier, to evaluate the three schemes, we implemented the baseline security protocols using OpenSSL library \cite{openssl} for CPUs and custom CUDA kernels on GPU. We were then able to make the necessary changes to this baseline to implement end-to-end training and inference pipelines with \name{}. 
We conducted all our experiments on a Nvidia H100 GPU. The CPU used in our setup is an Intel Xeon Gold 6548Y+ paired with 512GB of DRAM memory. Since the H100 GPU has CC capability, we were able to compare the end-to-end efficiency improvements of \name{} over a physical CC  implementation on Nvidia H100.

\subsection{Model/Data description} For benchmarking, we use three types of models: 1) the GraphSAGE graph convolution model~\cite{gcn}, (2) the ResNet50 vision model~\cite{resnet}, and 3) a Two-Tower neural network (TTNN) as the recommendation system~\cite{ttnn1, dlrm}. \textbf{GraphSAGE} analyzes graph-structured data through iterative neighborhood aggregation. We evaluated \name{} using a 3-layer GraphSAGE model~\cite{gcn} with a mean aggregator and neighborhood sample sizes of $S_1=S_2=S_3=10$ on the ogbn-products~\cite{hu2020ogb} dataset. During inference, users send node features and sub-graphs to predict product categories. \textbf{ResNet v1.5} is a variant of the original ResNet (Residual Network) architecture, specifically ResNet-50, modified to improve computational efficiency. We use the ResNet implementation from Nvidia's Deep Learning repositories~\cite{nvidia_deeplearning_examples}, with the ImageNet dataset~\cite{imagenet}. \textbf{TTNN} are commonly used in large-scale recommendation systems~\cite{ttnn1, dlrm}. These networks aim to compute relevance scores between a user and an item. For this task, we used the MovieLens 1M dataset~\cite{ml1m}.

\begin{figure}[htbp]
    \centering
    % Left subfigure (a)
    \begin{subfigure}{0.48\columnwidth}
        \centering
        \includegraphics[width=\linewidth, height=5.3cm]{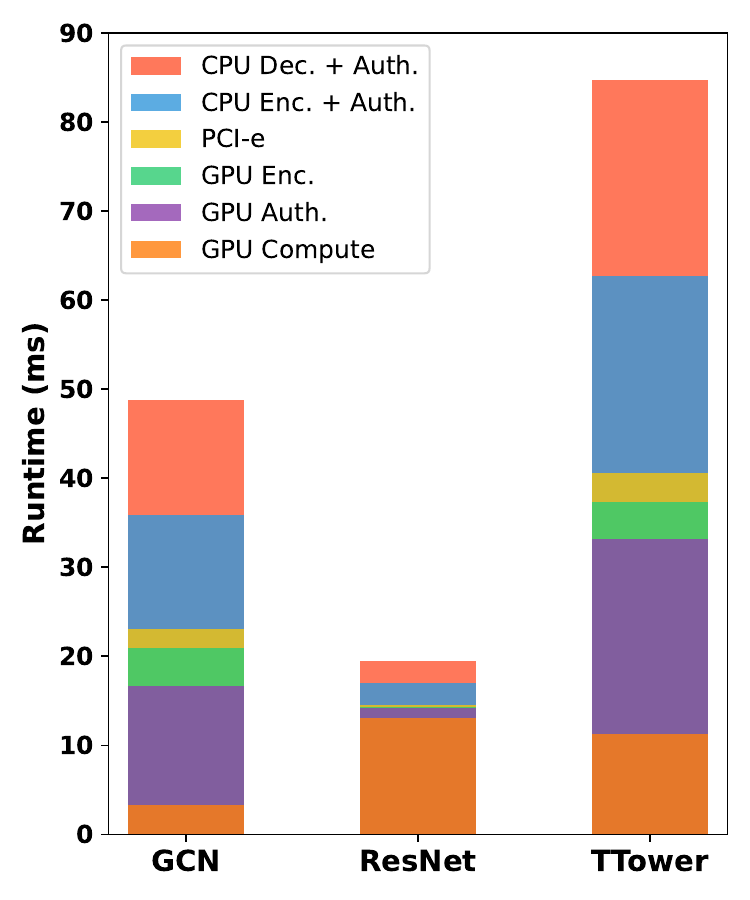}
        \caption{Inference.}
        \label{fig:infer-baseline}
    \end{subfigure}
    \hfill
    % Right subfigure (b)
    \begin{subfigure}{0.48\columnwidth}
        \centering
        \includegraphics[width=\linewidth, height=5.3cm]{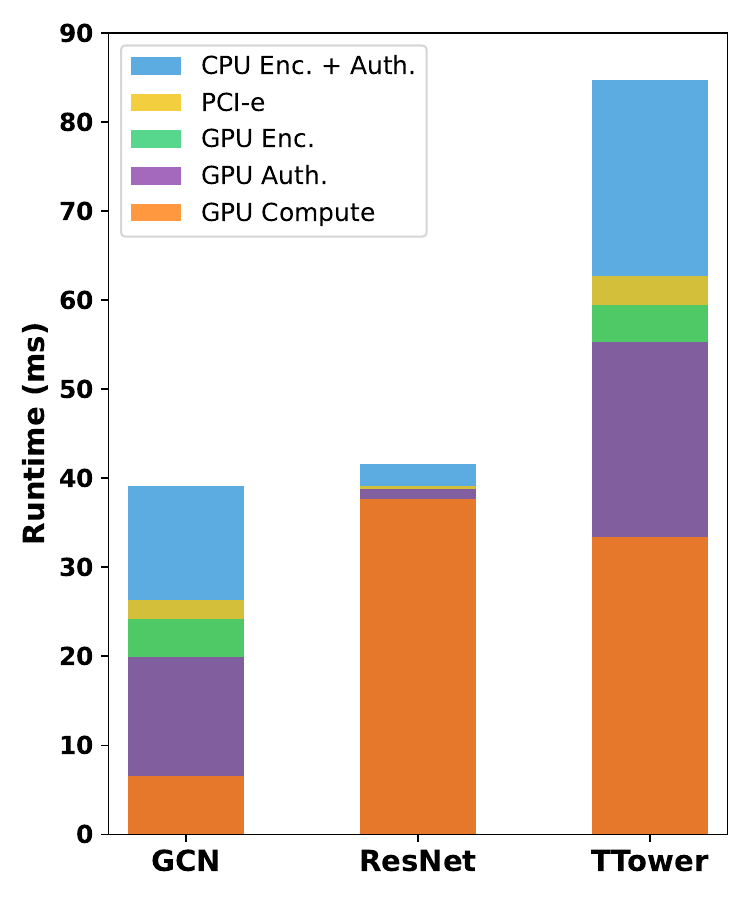}
        \caption{Training.}
        \label{fig:train-baseline}
    \end{subfigure}
    \caption{Baseline timing decomposition.}
    \label{fig:baseline-decomp}
\end{figure}

\begin{figure*}[htbp]
  \centering
  \begin{subfigure}[tb]{0.33\linewidth}
    \includegraphics[width=\linewidth, height=3.5cm]{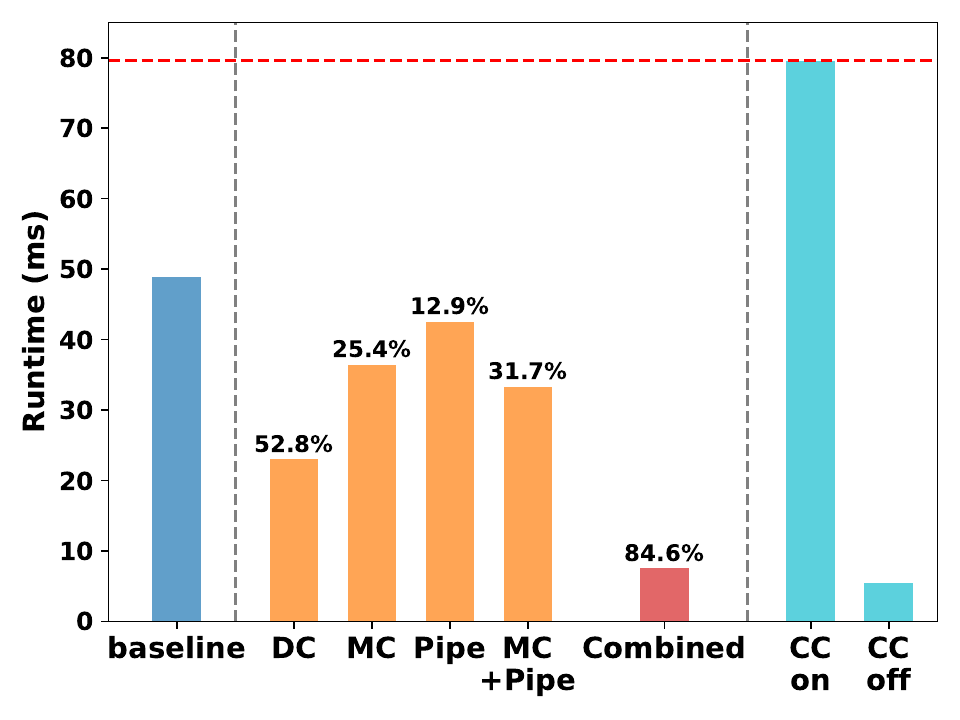}
    \caption{GCN Inference}
    \label{fig:gcn-infer}
  \end{subfigure}
  \begin{subfigure}[tb]{0.33\linewidth}
    \includegraphics[width=\linewidth, height=3.5cm]{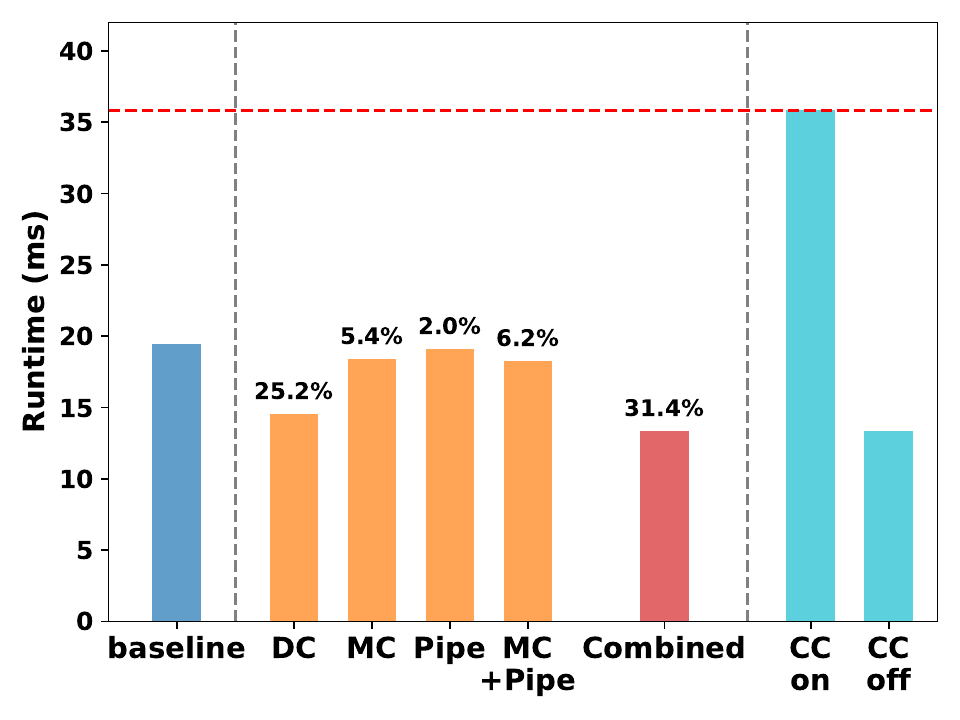}
    \caption{ResNet Inference}
    \label{fig:resnet-infer}
  \end{subfigure}
    \begin{subfigure}[tb]{0.33\linewidth}
    \includegraphics[width=\linewidth, height=3.5cm]{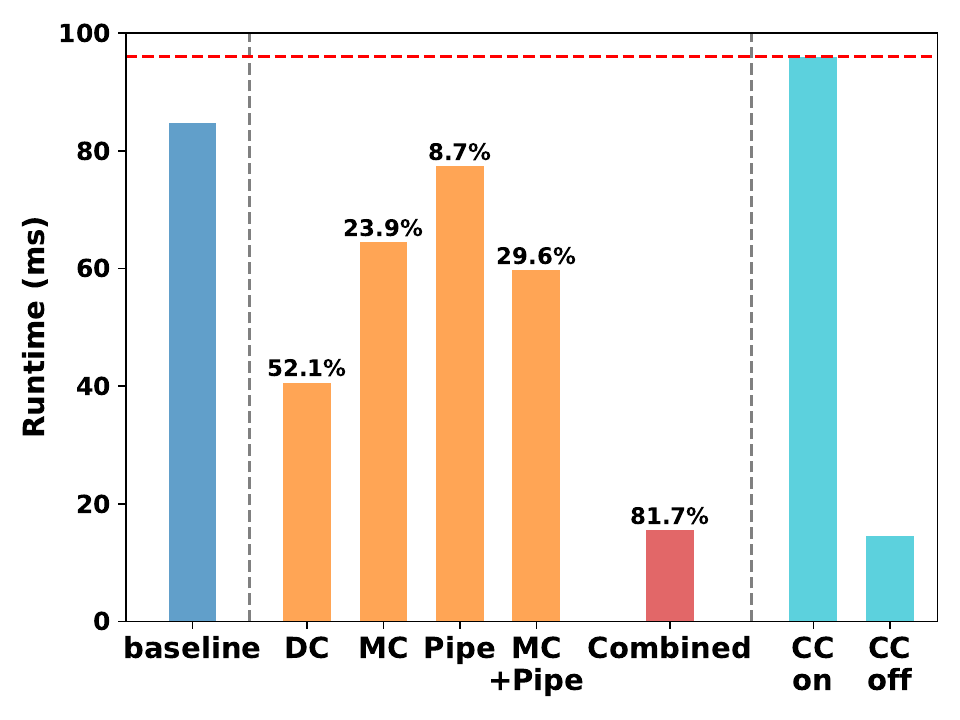}
    \caption{TTower Inference}
    \label{fig:tt-infer}
  \end{subfigure}

  \begin{subfigure}[tb]{0.33\linewidth}
    \includegraphics[width=\linewidth, height=3.5cm]{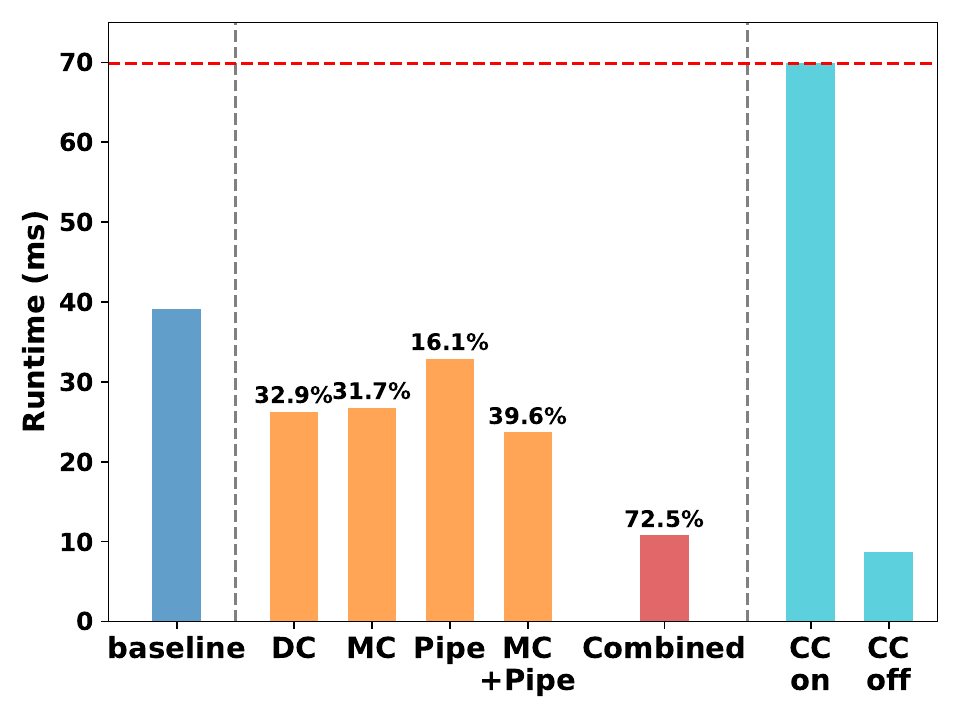}
    \caption{GCN Training}
    \label{fig:gcn-train}
  \end{subfigure}
  \begin{subfigure}[tb]{0.33\linewidth}
    \includegraphics[width=\linewidth, height=3.5cm]{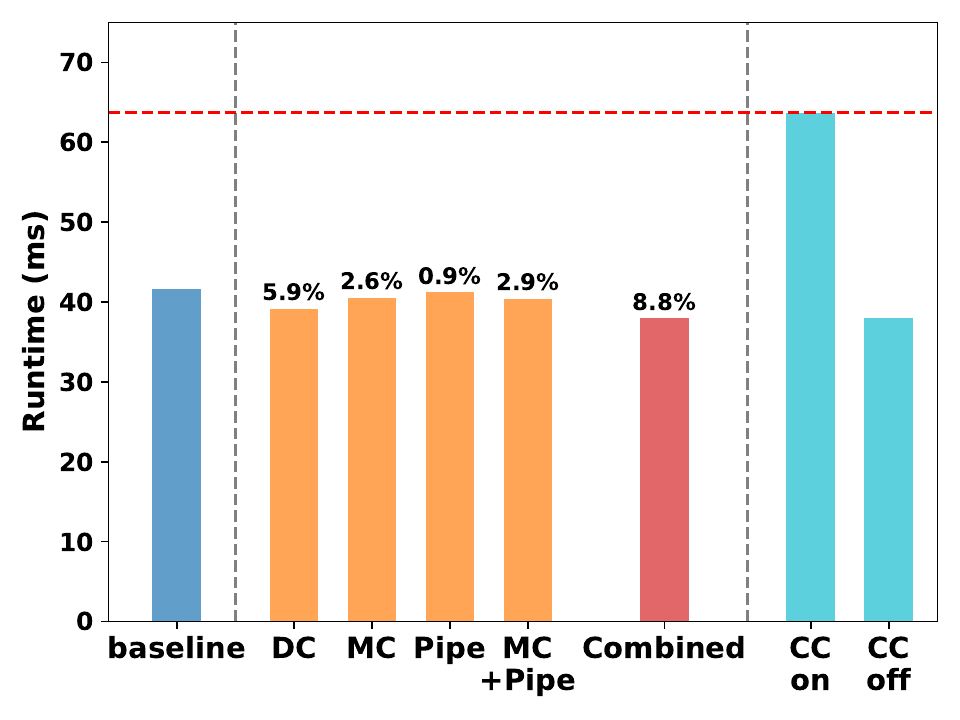}
    \caption{ResNet Training}
    \label{fig:resnet-train}
  \end{subfigure}
  \begin{subfigure}[tb]{0.33\linewidth}
    \includegraphics[width=\linewidth, height=3.5cm]{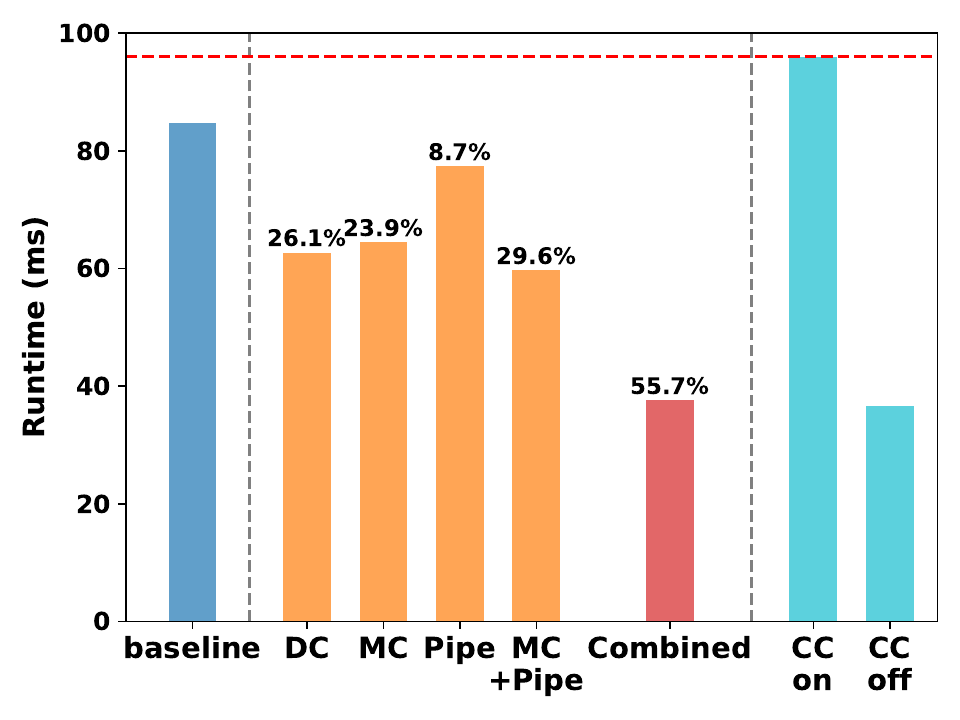}
    \caption{TTower Training}
    \label{fig:tt-train}
  \end{subfigure}
  \caption{\name{} runtime comparison vs Baseline, NVIDIA CC on and NVIDIA CC off}
  \label{fig:speedup}
\end{figure*}

\subsection{Baseline} Our baseline implementation for inference and training follow the flow shown in Figure~\ref{fig:baseline-compute}. Since we were able to implement the security protocols we can provide a detailed breakdown of the costs.  %This means that a penalty must be paid for data transfer due to CPU decryption (only for inference),  CPU encryption, CPU MAC generation, GPU decryption, and GPU authentication. An ideal system would incur no penalty for data transfer between the CPU and GPU TEE.
Figure~\ref{fig:baseline-decomp} shows the baseline runtime decomposition. For the ResNet50 model, the total runtime spent on securely transferring data is approximately 16.25 seconds per epoch, or 3.96 milliseconds per batch, with a batch size of 64. Of the 3.96 milliseconds per batch, around 312.5 microseconds are spent on actual PCI-e transfer, approximately 70.3 microseconds on GPU decryption, around 1.13 milliseconds on GPU authentication, and about 2.44 milliseconds on CPU encryption. This results in a 12.7$\times$ penalty (3.96ms vs. 312.5$\mu$s). For a full end-to-end transfer and compute pipeline, training with ResNet50 for a batch takes about 39.4 milliseconds, and inference takes about 17.36 milliseconds. ResNet50 spends approximately 10\% of its time during training on data transfer in GPU TEE mode and around 36.8\% of its time during inference with GPU TEE enabled.

During inference in a typical setting, the CPU TEE must first decrypt the user data, preprocess it (if necessary), then re-encrypt the data and generate a MAC before sending it to the GPU TEE. As AES encryption and decryption are essentially identical processes, the CPU incurs the penalty of performing encryption/decryption twice.
For GraphSAGE, the total end-to-end data communication time is around 6.23 seconds per epoch, or 32.6 milliseconds per batch, with a batch size of 1024. Of these 32.6 milliseconds on confidential data transfer, approximately 2.1 milliseconds are spent on data transfer, 12.98 milliseconds on CPU encryption and MAC generation, around 4.2 milliseconds on GPU decryption, and about 13.4 milliseconds on GPU authentication.

GPU authentication per batch is faster in GraphSAGE compared to ResNet50. Recall that the individual training samples (an image in Resnet versus a graph node in GraphSAGE) have very different sizes, and hence, the chain lengths required to compute an authentication tag differ. Overall, for GraphSAGE, we observe a slowdown of around 15.5X compared to encryption, decryption, and authentication-related work. Since computation in GraphSAGE is significantly faster than in ResNet50, the penalty for using GPU TEE is even larger for GCNs, as data transfer dominates execution time. As such, data transfer accounts for 83.24\% of GraphSAGE training time with GPU TEE, and this penalty is even greater during inference, where around 93.04\% of the total time is spent on data transfer with GPU TEE.

For the TTNN, the end-to-end data transfer/communication time is around 49.6 seconds per epoch, or 51.43 milliseconds per batch, with a batch size of 1024. Of the 51.43 milliseconds, approximately 2.19 milliseconds are spent on data transfer, 4.19 milliseconds on GPU decryption, 21.89 milliseconds on GPU authentication, and about 22.07 milliseconds on CPU encryption. For an end-to-end system using the TTNN, data transfer accounts for about 60.5\% of the total training time and around 86.6\% of the total inference time with GPU TEE enabled. In the subsequent subsections, we will describe the performance improvement of each optimization of \name{}. 

\subsection{Direct communication with GPU} 
We now present the advantages of direct communication with the GPU, i.e., removing the CPU-TEE from the critical path. Essentially, we change the computation workflow from Figure \ref{fig:baseline-compute} to Figure~\ref{fig:sharekey-compute}.
%As shown in Figure~\ref{fig:train-compute}, during training, the CPU TEEs fetch random batches of inputs and send them to the GPU TEEs. CPU must encrypt these inputs using the shared key between the CPU TEE and GPU TEE and generate MAC tags for authentication. \name{} eliminates all those CPU computations from the critical path, thereby removing the need to pay the penalty for encrypting and MAC tag generation before PCI-e transfer. For inference, the CPU TEE must perform an additional decryption step on the user data before re-encrypting it, further prolonging the data processing/transfer time, as seen in Figure \ref{fig:infer-compute}. 
Figure \ref{fig:speedup} illustrates the runtime reduction achieved through our optimizations (the number on top of each bar represents the percentage runtime reduction for each optimization) for an end-to-end system during training and inference. The baseline execution time is represented by the first bar, shown in blue. The runtime reduction due to direct communication is displayed in the second column of each of the six figures, marked with "DC" (direct communication).
As observed, communication-intensive workloads like GCN and TTNN experience significant runtime reduction. In all cases, the runtime reduction for inference is even better than for training. %As mentioned earlier, this is because, during inference, the CPU TEE must first decrypt the data, preprocess it (if necessary), and then re-encrypt it before sending it to the GPU. Essentially, the penalty for CPU encryption is doubled during inference, so eliminating this step results in a significant overall performance boost. 
The runtime reduction for ResNet training is the lowest, 5.9\%. This is because, even with GPU TEE, data transfer only accounts for about 10\% of the overall training time. Therefore, the maximum possible runtime reduction for ResNet training with GPU TEE is around 10\%. In contrast, for TTNN and GraphSAGE inference, we achieve up to a ~53\% runtime reduction by removing CPU from the critical path.

\subsection{Multi-chaining authentication} %Our primary objective with \name{} is to make the overhead of using GPU TEEs negligible. GPU authentication is a particularly slow process (the slowest security process for the GPU), as it is a sequential algorithm, as discussed in \ref{subsec:aesgcm}. In our baseline, each input sample is assigned a single MAC tag, introducing some level of parallelism. Despite this, the baseline still spends a considerable amount of time on authentication.

We now measure the impact of proposed multi-chaining authentication where each input sample is broken into multiple chunks/chains, and each chunk will have an associated MAC tag. In our experiments, we used 16 tags per item in each batch. This optimization reduces the GPU authentication time by up to 16$\times$ since the chain length for AES-GCM is now 16 times shorter. The third bar in Figure \ref{fig:speedup} illustrates the runtime reduction for \name{} when using 16 tags per item. With this multi-chaining optimization, we achieve up to a 31.7\% reduction in runtime. ResNet does not show significant performance improvements because authentication costs are smaller fraction of the total latency. %data transfer already occupies a small fraction of the overall time.

There are two key points of interest here. First, reducing authentication time is not only about speeding up the end-to-end system but also about ensuring that the authentication time is shorter than the actual data transfer time. This becomes crucial when we implement decryption pipeline optimizations, allowing us to pipeline and fully hide the GPU authentication behind data transfer. Second, although using multiple tags for each batch element increases the total data sent to the GPU (since all these additional tags must be transferred for authentication checks), the overall increase in data transfer size is only about 0.5\%. This increase causes no noticeable slowdowns in other parts of the process.

\subsection{Decryption pipeline} Our final optimization for \name{} is to pipeline data transfer, GPU decryption, and GPU authentication. Since the GPU authentication kernel utilizes only 5$\%$ of the GPU resources, the GPU can execute both kernels completely in parallel, completing those two kernels with the runtime of the longer one. This effectively reduces the overall transfer time to the maximum of the data transfer, GPU authentication, and GPU decryption times, amortized over an entire epoch. The fourth bar in Figure \ref{fig:speedup} shows this runtime reduction.

The fifth bar in each figure of Figure \ref{fig:speedup} shows the total runtime reduction when applying multi-chaining and the decryption pipeline together. We observe up to a 39.6\% runtime reduction by using both optimizations together. Notably, the runtime reduction for TTNN training and inference are identical for multi-chaining, pipelining, and the combined result. This is because the baseline execution times for both are the same. The additional backward pass in TTNN training takes about the same amount of time as the extra CPU decryption time in TTNN inference.

The sixth/red bar represents runtime reduction when combining all optimizations. We achieve up to an 84.6\% runtime reduction (an improvement of 6.49$\times$) over the baseline for end-to-end inference and up to a 72.5\% runtime reduction (an improvement of 3.63$\times$) for end-to-end training with GraphSAGE. For ResNet50, we achieve up to a 31.4\% runtime reduction for end-to-end inference and up to an 8.8\% runtime reduction for end-to-end training. For the TTNN, we achieve up to an 81.7\% runtime reduction (an improvement of 5.46$\times$) for end-to-end inference and up to a 55.7\% runtime reduction (an improvement of 2.26$\times$) for end-to-end training. GraphSAGE has the biggest runtime gap between \name{} and GPUs without CC. This is because, even without GPU CC, data transfer remains a significant portion of the total execution time, and \name{} was not able to fully hide GPU decryption behind data transfer. Although GPU authentication could be hidden behind transfer due to multi-chaining, GPU decryption could not be hidden. For GraphSAGE inference, \name{} experienced a 38.7\% slowdown compared to execution without GPU CC, and training experienced a 24.2\% slowdown.

%YR: ADD THIS SECTION WHEN READY
%\subsection{Comparison with GPUs CC} The final bar, shown in light blue, represents the runtime of the non-GPU TEE version. For ResNet50, we were able to completely hide GPU authentication and GPU decryption under the data transfer time, so \name{} performs as well as the non-GPU TEE version. For the TTNN, inference incurs a 6.9\% slowdown with \name{} compared to the non-GPU TEE version, while training results in a 2.7\% slowdown with \name{}.

\subsection{Comparison with Physical Hardware}
The last two bars represent the runtime of physical GPU hardware with ("CC on") and without CC ("CC off"). We were able to turn on CC on an H100 GPU, and hence, we can provide end-to-end latency measurements, even though we can't provide a detailed breakdown.

\para{Comparison with CC}
Now, we compare the end-to-end performance of \name{} with the physical hardware implementation of H100 GPU with NVIDIA CC enabled. Note that, as stated earlier, since the implementation details of H100 CC are unavailable, we only provide the observed times as measured on physical hardware. 
Across the board, the results show that \name{} (6th bar) can outperform GPU CC on real physical hardware for all models and sometimes even comes close to the execution time of H100 without CC, where the workload is executed without any security protection. We describe that comparison next. The last bar in each part of Figure \ref{fig:speedup} shows execution time for CC off. 

\para{Comparison without CC}
Comparing the end-to-end latency on H100 without CC with \name{} we show that for ResNet50, we were able to completely hide GPU authentication and GPU decryption under the data transfer time, so \name{} performs as well as the GPU without CC. For the TTNN, inference incurs a 6.9\% slowdown with \name{} compared to GPU without CC, while training results in a 2.7\% slowdown with \name{}. Those slowdowns are significantly smaller than Nvidia systems with CC on. The seventh bar in each part of figure \ref{fig:speedup} shows the execution time for CC off.

\subsection{Large language models (LLMs)} Since we only had access to one H100 GPU, LLM training evaluations are impractical.   LLMs are typically trained in model parallel distributed environments, where the overhead is particularly significant due to the need to transfer activation maps and gradients across GPUs. We can only surmise that such a system may provide opportunities for \name{} to work well. However, due to computing limitations, we do not have data from distributed training setups to fully evaluate these potential gains.

\section{Related Works}
\label{sec:related}
\name{} utilizes Intel TDX’s confidential VM (CVM) to manage data movement within CPU TEEs. While CPU-based TEEs, such as Intel SGX~\cite{sgx}, ARM TrustZone~\cite{trustzon}, Intel TDX~\cite{tdx}, and AMD SEV~\cite{amdcc}, provide enhanced security, their limited parallelism has led to leveraging untrusted GPUs for computationally heavy ML tasks~\cite{tramer2019slalom, darkNight, origami, li2024translinkguard}. In contrast, \name{} uses a secure GPU with TEE to run all stages of ML model training and inference within a fully trusted environment, ensuring data integrity and confidentiality while benefiting from GPU parallelism for better performance.

\para{GPU-based TEE works} Graviton~\cite{graviton} is the first to propose GPU-based TEE, and HIX~\cite{hix} proposed a GPU-based TEE systems without modifying GPU implementations. Nvidia H100 is the first commercially available GPU with TEE functionalities (Nvidia confidential computing). \name{} uses Nvidia H100 as the baseline and demonstrate \name{}'s performance improvements. \cite{distcc} focuses on dynamically managing AES pads for distributed training, while \name{} focuses on the CPU-to-GPU IO overheads during single-node ML model inference and training.

% \para{Other privacy-preserving techniques} There exists other privacy-preserving technologies to enhance data security in the cloud, such as differential privacy, homomorphic encryption, and Secure multi-party computation. Those techniques have different threat models than GPU TEE systems and are not the focus of our paper.

\section{Conclusion}
\label{sec:conclusion}
In this paper, we propose \name{} to reduce CPU-to-GPU I/O overheads in GPU TEE systems. We identified three inefficiencies: 1) CPU re-encryption, 2) lack of parallelism in authentication, and 3) operation serialization. To address these, we introduced a direct communication link with GPU TEEs, multi-chaining authentication, and a decryption pipeline. These techniques significantly improve ML performance in GPU TEE systems while maintaining the same security level as Nvidia CC.

\bibliographystyle{plain}
\bibliography{references}

\begin{thebibliography}{10}

\bibitem{snoop1}
I2c bus monitor.
\newblock \url{https://www.jupiteri.com/}.
\newblock Accessed: 2023-08-3.

\bibitem{amdcc}
AMD.
\newblock Amd secure encrypted virtualization (sev).
\newblock \url{https://www.amd.com/en/developer/sev.html}.
\newblock Accessed: 2024-10-07.

\bibitem{biasttnnpaper}
Keshav Balasubramanian, Abdulla Alshabanah, Elan Markowitz, Greg Ver~Steeg, and Murali Annavaram.
\newblock Biased user history synthesis for personalized long-tail item recommendation.
\newblock In {\em Proceedings of the 18th ACM Conference on Recommender Systems}, RecSys ’24, page 189–199, New York, NY, USA, 2024. Association for Computing Machinery.

\bibitem{imagenet}
Jia Deng, Wei Dong, Richard Socher, Li{-}Jia Li, Kai Li, and Li~Fei-Fei.
\newblock Imagenet: A large-scale hierarchical image database.
\newblock In {\em 2009 IEEE Conference on Computer Vision and Pattern Recognition (CVPR)}, pages 248--255, 2009.

\bibitem{aesgcm}
Morris Dworkin.
\newblock Recommendation for block cipher modes of operation: Galois/counter mode (gcm) and gmac.
\newblock {\em NIST Special Publication 800-38D}, 2007.

\bibitem{gcn}
Will Hamilton, Zhitao Ying, and Jure Leskovec.
\newblock Inductive representation learning on large graphs.
\newblock {\em Advances in neural information processing systems}, 30, 2017.

\bibitem{ml1m}
F.~Maxwell Harper and Joseph~A. Konstan.
\newblock The movielens datasets: History and context.
\newblock \url{https://grouplens.org/datasets/movielens/1m/}, 2016.
\newblock ACM Transactions on Interactive Intelligent Systems (TiiS), 5(4), 2016. Accessed: 2024-10-18.

\bibitem{darkNight}
Hanieh Hashemi, Yongqin Wang, and Murali Annavaram.
\newblock Darknight: A data privacy scheme for training and inference of deep neural networks.
\newblock {\em Proceedings on the 54th International Symposium on Microarchitecture}, 2021.

\bibitem{hashemi2022data}
Hanieh Hashemi, Wenjie Xiong, Liu Ke, Kiwan Maeng, Murali Annavaram, G.~Edward Suh, and Hsien-Hsin~S. Lee.
\newblock Data leakage via access patterns of sparse features in deep learning-based recommendation systems.
\newblock 2022.

\bibitem{resnet}
Kaiming He, Xiangyu Zhang, Shaoqing Ren, and Jian Sun.
\newblock Deep residual learning for image recognition.
\newblock {\em CoRR}, abs/1512.03385, 2015.

\bibitem{hu2020ogb}
Weihua Hu, Matthias Fey, Marinka Zitnik, Yuxiao Dong, Hongyu Ren, Bowen Liu, Michele Catasta, and Jure Leskovec.
\newblock Open graph benchmark: Datasets for machine learning on graphs.
\newblock In {\em Proceedings of the 34th Conference on Neural Information Processing Systems (NeurIPS)}, 2020.

\bibitem{tdx}
Intel.
\newblock Intel trust domain extensions.
\newblock \url{https://www.intel.com/content/www/us/en/developer/tools/trust-domain-extensions/overview.html}.
\newblock Accessed: 2024-10-07.

\bibitem{hix}
Insu Jang, Adrian Tang, Taehoon Kim, Simha Sethumadhavan, and Jaehyuk Huh.
\newblock Heterogeneous isolated execution for commodity gpus.
\newblock In {\em Proceedings of the Twenty-Fourth International Conference on Architectural Support for Programming Languages and Operating Systems}, ASPLOS '19, page 455–468, New York, NY, USA, 2019. Association for Computing Machinery.

\bibitem{sgx}
Simon Johnson, Vinnie Scarlata, Carlos Rozas, Ernie Brickell, and Frank Mckeen.
\newblock Intel software guard extensions: Epid provisioning and attestation services.
\newblock 2016.

\bibitem{li2024translinkguard}
Qinfeng Li, Zhiqiang Shen, Zhenghan Qin, Yangfan Xie, Xuhong Zhang, Tianyu Du, Sheng Cheng, Xun Wang, and Jianwei Yin.
\newblock Translinkguard: Safeguarding transformer models against model stealing in edge deployment.
\newblock In {\em ACM Multimedia 2024}, 2024.

\bibitem{trustzon}
ARM Limted.
\newblock Arm security technology building a secure system using trustzone technology.
\newblock 2016.

\bibitem{distcc}
Seonjin Na, Jungwoo Kim, Sunho Lee, and Jaehyuk Huh.
\newblock Supporting secure multi-gpu computing with dynamic and batched metadata management.
\newblock In {\em 2024 IEEE International Symposium on High-Performance Computer Architecture (HPCA)}, pages 204--217, 2024.

\bibitem{origami}
Krishna~Giri Narra, Zhifeng Lin, Yongqin Wang, Keshav Balasubramaniam, and Murali Annavaram.
\newblock Privacy-preserving inference in machine learning services using trusted execution environments.
\newblock {\em IEEE International Conference on Cloud Computing}, 2021.

\bibitem{dlrm}
Maxim Naumov, Dheevatsa Mudigere, Hao{-}Jun~Michael Shi, Jianyu Huang, Narayanan Sundaraman, Jongsoo Park, Xiaodong Wang, Udit Gupta, Carole{-}Jean Wu, Alisson~G. Azzolini, Dmytro Dzhulgakov, Andrey Mallevich, Ilia Cherniavskii, Yinghai Lu, Raghuraman Krishnamoorthi, Ansha Yu, Volodymyr Kondratenko, Stephanie Pereira, Xianjie Chen, Wenlin Chen, Vijay Rao, Bill Jia, Liang Xiong, and Misha Smelyanskiy.
\newblock Deep learning recommendation model for personalization and recommendation systems.
\newblock {\em CoRR}, abs/1906.00091, 2019.

\bibitem{nvdiacc}
NVIDIA.
\newblock Nvidia confidential computing.
\newblock \url{https://www.nvidia.com/en-us/data-center/solutions/confidential-computing/}.
\newblock Accessed: 2024-10-07.

\bibitem{h100}
NVIDIA.
\newblock Nvidia h100 tensor core gpu: Unprecedented performance, scalability, and security.
\newblock \url{https://www.nvidia.com/en-us/data-center/h100/}, 2022.
\newblock Accessed: 2024-10-18.

\bibitem{blackwell}
NVIDIA.
\newblock Nvidia blackwell architecture.
\newblock \url{https://www.nvidia.com/blackwell/}, 2024.
\newblock Accessed: 2024-10-18.

\bibitem{dali}
NVIDIA.
\newblock Nvidia dali: The nvidia data loading library.
\newblock \url{https://github.com/NVIDIA/DALI}, 2024.
\newblock Accessed: 2024-10-18.

\bibitem{nvidia_deeplearning_examples}
NVIDIA.
\newblock Nvidia deep learning examples.
\newblock \url{https://github.com/NVIDIA/DeepLearningExamples}, 2024.
\newblock Accessed: 2024-10-18.

\bibitem{openssl}
OpenSSL Project.
\newblock Openssl: The open source toolkit for ssl/tls.
\newblock \url{https://www.openssl.org/}, 2024.
\newblock Accessed: 2024-10-18.

\bibitem{tramer2019slalom}
Florian Tramèr and Dan Boneh.
\newblock Slalom: Fast, verifiable and private execution of neural networks in trusted hardware.
\newblock {\em arXiv preprint arXiv:1806.03287}, 2019.

\bibitem{graviton}
Stavros Volos, Kapil Vaswani, and Rodrigo Bruno.
\newblock Graviton: Trusted execution environments on {GPUs}.
\newblock In {\em 13th USENIX Symposium on Operating Systems Design and Implementation (OSDI 18)}, pages 681--696, Carlsbad, CA, October 2018. USENIX Association.

\bibitem{jpeg}
Gregory~K. Wallace.
\newblock Jpeg: Still image compression standard, 1992.

\bibitem{ttnn1}
Xinyang Yi, Ji~Yang, Lichan Hong, Derek~Zhiyuan Cheng, Lukasz Heldt, Aditee~Ajit Kumthekar, Zhe Zhao, Li~Wei, and Ed~Chi, editors.
\newblock {\em Sampling-Bias-Corrected Neural Modeling for Large Corpus Item Recommendations}, 2019.

\end{thebibliography}

\end{document}